\documentclass[english,showpacs,prb,twocolumn]{revtex4}

\usepackage{graphicx}
\usepackage{enumerate}
\usepackage{hyperref}
\usepackage{textcomp}
\usepackage{amsmath}
\usepackage{amssymb}
\usepackage{pgf}
\usepackage{subfigure}
\usepackage[english]{babel}
\usepackage {cancel}

\def\ud{\mathrm{d}}

\newcommand{\nep}{\textrm{e}}

\newcommand{\LZ}{\mathrm{\scriptscriptstyle LZ}}
\newcommand{\res}{\mathrm{\scriptstyle res}}

\newcommand{\calH}{\mathcal{H}}

\newcommand{\opS}{\widehat{S}}

\newcommand{\prm}{p}
\newcommand{\rmf}{\mathrm{f}}
\newcommand{\rmi}{\mathrm{i}}
\newcommand{\Pm}{\mathbb P}
\newcommand{\PauliSigma}{\hat{\sigma}}
\newcommand{\sigmabar}{\overline{\sigma}}

\begin{document}

\title{Direct comparison of quantum and simulated annealing on a fully-connected Ising ferromagnet}

\author{Matteo M. Wauters$^{1}$, Rosario Fazio$^{2,3}$, Hidetoshi Nishimori$^4$, Giuseppe E. Santoro$^{1,2,5}$}

\affiliation{
$^1$ SISSA, Via Bonomea 265, I-34136 Trieste, Italy\\
$^2$ International Centre for Theoretical Physics (ICTP), Strada Costiera 11, I-34014 Trieste, Italy\\
$^3$ NEST, Scuola Normale Superiore and Istituto Nanoscienze-CNR, I-56126 Pisa, Italy\\
$^4$ Dept. of Physics, Tokyo Institute of Technology, Tokyo 152-8551, Japan \\
$^5$ CNR-IOM Democritos National Simulation Center, Via Bonomea 265, I-34136 Trieste, Italy
}

\begin{abstract}
We compare the performance of quantum annealing (QA, through Schr\"odinger dynamics) and simulated annealing 
(SA, through a classical master equation) on the $p$-spin infinite range ferromagnetic Ising model, 
by slowly driving the system across its equilibrium, quantum or classical, phase transition. 
When the phase transition is second-order ($\prm=2$, the familiar two-spin Ising interaction) SA shows a remarkable 
exponential speed-up over QA.
For a first-order phase transition ($\prm\ge 3$, i.e., with multi-spin Ising interactions) 
, in contrast, the classical annealing 
dynamics appears to remain stuck in the disordered phase, while we have clear evidence that QA shows a residual energy which decreases towards $0$ when the total annealing time $\tau$ increases, albeit in a rather slow (logarithmic) fashion.  This is one of the rare examples where a limited quantum speedup, a speedup by QA over SA, has been shown to exist by direct solutions of the Schr\"odinger and master equations in combination with a non-equilibrium Landau-Zener analysis.
We also analyse the imaginary-time QA dynamics of the model, finding a $1/\tau^2$ behaviour for all finite values of $\prm$, 
as predicted by the adiabatic theorem of quantum mechanics. The Grover-search limit $p({\rm odd})=\infty$ is also discussed. 
\end{abstract}

\maketitle
\section{Introduction} \label{intro:sec}
Many of the complex problems of interest, notably all combinatorial optimization problems, can be generally cast as
the search for the global minimum energy state of a suitable classical Ising Hamiltonian, depending on $N$ binary (spin) variables 
\cite{Lucas2014}.
Quantum annealing (QA) \cite{Finnila_CPL94,Kadowaki_PRE98,Brooke_SCI99,Santoro_SCI02,Santoro_JPA06,Morita_JMP08}
--- intimately related to Adiabatic Quantum Computation  \cite{Farhi_SCI01} ---
was originally thought as an alternative route to classical annealing \cite{Kirkpatrick_SCI83}, employing quantum fluctuations 
to effectively escape, through quantum tunnelling, from unfavourable local minima in a complex energy landscape. 

QA has lately become a very active field of research, due to the availability of quantum annealing 
programmable machines based on superconducting flux quantum bits \cite{Harris_PRB10,Johnson_Nat11}.
One of the open issues in the field is finding classes of problems where  clear evidence is seen for a {\em limited }{\em quantum speedup} in the sense defined in Ref.\onlinecite{Ronnow2014},
i.e., a QA based computation would show a better scaling with the number of variables $N$ than a corresponding classical heuristics,
for instance Simulated Annealing (SA) \cite{Kirkpatrick_SCI83}. 
The original positive results on the random Ising model \cite{Santoro_SCI02,Martonak_PRB02}, followed by equally encouraging
results on the Travelling Salesman Problem \cite{Martonak_PRE04}, were soon followed by a rather disappointing outcome on
a random Satisfiability problem \cite{Battaglia_PRE05}. 
By now, the literature on this issue has grown enormously. A partial list is Refs. 
\onlinecite{Stella_PRB05,Stella_PRB06,Matsuda2009,Young2010,Hen2011,Farhi2012,Boixo2014,Katzgraber2014,Katzgraber2015,
Ronnow2014,Albash2015,Heim2015,Hen2015,Isakov2015,Martin-mayor2015,Liu_2015,Steiger2015,Venturelli2015,
Crosson2016,Denchev2016,Kechedzhi2016,Mandra2016,Mandra2016b,Marshall2016,Muthukrishnan2016,Isakov2016,AlbashLidar2016,King2017,Mandra2017,AlbashLidar2017,Jiang2017,Herr2017} and the references therein cited.

It is fair to say that the picture is far from complete. 
One of the notable exceptions where a quantum speedup has been clearly proven is Grover's search problem\cite{Grover_PRL97,Roland_PRA02} 
--- searching for a given item in a unsorted database of $\mathcal{N}=2^N$ items --- where a quadratic speedup, from 
the classical algorithm scaling as $\sim 2^N$ to a quantum algorithm scaling as $2^{N/2}$ is known \cite{Grover_PRL97,Roland_PRA02}. 

The Grover problem can be regarded as a particular limit \cite{Jorg_EPL10} of a fully-connected Ising model with 
uniform ferromagnetic couplings, whose classical Hamiltonian reads:
\begin{equation} \label{H_C:eqn}
H_{\rm C}= -\frac{JN}{2}\left(\frac{1}{N}\sum_{j=1}^N \PauliSigma_j^z \right)^{\prm} \;,
\end{equation}
where $\prm\ge 2$ is an integer parameter, and we have expressed the classical spin variables in terms of 
$\PauliSigma^z$ Pauli matrices. 
In the limit in which one sets $\prm (\mathrm{odd}) = \infty$ the model effectively describes the Grover problem: 
a single isolated minimal energy configuration, the fully magnetized state $|\!\uparrow,\uparrow,\dots,\uparrow\rangle$
has energy $-JN/2$, and is well separated from all other configurations having zero energy. 

\begin{table*}[ht]
\begin{tabular}{| c | c | c  | c | c | c | c |}
\hline
\multicolumn{7}{| c |}{}\\
\multicolumn{7}{| c |}{{\bf Annealing of the $p$-spin fully connected Ising model}}\\
\multicolumn{7}{| c |}{}\\
\multicolumn{2}{| l }{QA (Schr\"odinger) dynamics:} & 
\multicolumn{4}{l}{${\widehat H}_{\rm Q}(t)=-\frac{JN}{2}\left(\frac{1}{N}\sum_{j=1}^N \PauliSigma^z_j \right)^{\prm} - 
\Gamma(t)\sum_{i=1}^N \PauliSigma _i^x$} &
\multicolumn{1}{l | }{$\Gamma(t)=\Gamma_{\rmi}(1-\frac{t}{\tau})+\Gamma_{\rmf} \frac{t}{\tau}$} \\
\multicolumn{7}{| l |}{}\\
\multicolumn{2}{| l }{SA (Master eq.) dynamics:} & 
\multicolumn{4}{l}{${H}_{\rm C}=-\frac{JN}{2}\left(\frac{1}{N}\sum_{j=1}^N \sigma_j \right)^{\prm}$} & 
\multicolumn{1}{l |}{$T(t)=T_{\rmi}(1-\frac{t}{\tau}) + T_{\rmf} \frac{t}{\tau}$}\\
\multicolumn{7}{| l |}{}\\
\hline
\hline
& \multicolumn{4}{|c|}{$\epsilon^{\res}_N(\tau)$} 
& \multicolumn{2}{|c|}{$\epsilon^{\res}_{N\to \infty}(\tau)$} \\
\cline{2-7} 
& \multicolumn{2}{|c|}{$\prm=2$} & \multicolumn{2}{|c|}{$\prm\ge 3$}
& \multicolumn{1}{|c|}{$\prm=2$} & \multicolumn{1}{|c|}{$\prm\ge 3$} \\
& intermediate $\tau$ & asymptotic & intermediate $\tau$ & asymptotic &
asymptotic & asymptotic \\
\hline
\hline
 & & & & & & \\
QA-RT & $\frac{C_p}{N}\nep^{-\tau/\tau^*_N}$ & ${\displaystyle \frac{\Gamma_{\rmi}^2}{8} \frac{1}{\tau^2}}$ \footnotemark
\footnotetext{ These results follow from the adiabatic theorem of Ref.~\onlinecite{Morita_JMP08} } &
$\frac{C_p}{N}\nep^{-\tau/\tau^*_{N,\prm}}$   & ${\displaystyle \frac{\Gamma_{\rmi}^2}{p^3} \frac{1}{\tau^2}}$ & \hspace{0mm} 
${\displaystyle \sim \frac{1}{\tau^{3/2}}}$ \footnotemark \footnotetext{ Previously found in Ref.~\onlinecite{Caneva_PRB08}}  
\hspace{0mm} & \hspace{0mm} ${\displaystyle \sim \frac{1}{log(\gamma \tau)}}$ \hspace{0mm}
\\
            & ${\scriptstyle (\tau_N^*\sim N^{2/3})}$  &  & ${\scriptstyle (\tau_{N,\prm}^*\sim \nep^{2\alpha_p N})}$ 
            \footnotemark \footnotetext{ Follows also from the critical gap scaling found in Refs.~\onlinecite{Jorg_EPL10,Bapst_JSTAT12} }& 
            & {\scriptsize (See Eq.~\eqref{eqn:env_p2})}& {\scriptsize (See Eq.~\eqref{eqn:eres_qa_rt_p3})} \\
\hline
 & \multicolumn{2}{|c|}{} & \multicolumn{2}{|c|}{} & & \\
QA-IT & \multicolumn{2}{|c|}{${\displaystyle \frac{\Gamma_{\rmi}^2}{8} \frac{1}{\tau^2}}$} &  \multicolumn{2}{|c|}{${\displaystyle \frac{\Gamma_{\rmi}^2}{p^3} \frac{1}{\tau^2}}$} 
&${\displaystyle \frac{\Gamma_{\rmi}^2}{8} \frac{1}{\tau^2}}$ & ${\displaystyle \frac{\Gamma_{\rmi}^2}{p^3} \frac{1}{\tau^2}}$ \\
 & \multicolumn{2}{|c|}{} & \multicolumn{2}{|c|}{} & & \\
\hline
\hline
 & & & & & & \\
SA ($T_{\rm f}>0$) & $\frac{1}{2} \nep^{-\tau/\tau_{\rm f}'}$ & ${\displaystyle \frac{C}{\tau}}$ & $\frac{1}{2}\nep^{-\tau/\tau^*_{N,\prm,{\rm f}}}$  
& ${\displaystyle \frac{C}{\tau}}$ & ${\displaystyle \frac{C}{\tau}}$ & ${\displaystyle \frac{1}{2}}$ \\
 & &  & ${\scriptstyle (\tau_{N,\prm,{\rm f}}^*\sim  \nep^{N A_{\prm,T_{\rm f}}})}$ & & & \\
 & & & & & & \\
\hline
 & & & & & & \\
SA ($T_{\rm f}=0$) & - & $\sim \nep^{-\tau/\tau_{\rm f}}$ & - & $\frac{1}{2}\nep^{-\tau/\tau^*_{N,\prm}}$ & $\sim \nep^{-\tau/\tau_{\rm f}}$ & 
${\displaystyle \frac{1}{2}}$ \\
 & &  ${\scriptstyle (N-{\rm indep.} \tau_{\rm f})}$ & & ${\scriptstyle (\tau_{N,\prm}^*\sim  N^{\frac{\prm-2}{2}})}$ &${\scriptstyle (N-{\rm indep.} \tau_{\rm f})}$& \\
 & & & & & & \\
\hline
\end{tabular}
\caption{Summary of our findings for the behaviour of the residual energy $\epsilon^{\res}_N(\tau)$ at the end of a linear annealing 
over a time $\tau$ at finite $N$ and in the thermodynamic limit $\epsilon^{\res}_{N\to \infty}(\tau)$. 
For a definition of the $\epsilon^{\res}_N(\tau)$, see Eq.~\eqref{eqn:resQC}.}
\label{table:summary}
\end{table*}

Even for the problem defined by the Hamiltonian in Eq. (\ref{H_C:eqn}), a definite answer for generic $p$ is still missing for dynamics, although statistical-mechanical analyses lead to a good amount of understanding of equilibrium properties.\cite{Jorg_EPL10,Bapst_JSTAT12,NishimoriTakada2017}
It is interesting to ask whether and when a quantum speedup can be proven in this particular class of simple problems. 
Studies on the QA of the fully-connected Ising ferromagnet have appeared in Refs.~\onlinecite{Caneva_PRB08,Jorg_EPL10,Bapst_JSTAT12}. 
The common setting is to supplement the classical ``potential energy'' $H_{\rm C}$ with a transverse field term 
$-\Gamma(t)\sum_j \PauliSigma^x_j$, obtaining a time-dependent quantum Hamiltonian of the form
\begin{equation}\label{eq:H_Q}
{\widehat H}_{\rm Q}(t)=-\frac{JN}{2}\left(\frac{1}{N}\sum_{j=1}^N \PauliSigma^z_j \right)^{\prm} 
- \Gamma(t)\sum_{i=1}^N \PauliSigma _i^x \;, 
\end{equation}
whose Schr\"odinger unitary evolution can be studied, exploiting permutation symmetry, for rather large values of $N\sim 1000$.
The efficiency of QA is remarkably related to the nature of the transition point separating the large-$\Gamma$ 
quantum paramagnetic phase from the small-$\Gamma$ ferromagnetic phase. 
It is known that for $\prm=2$ the transition is of second-order, and the resulting QA evolution improves its 
estimate of the ground-state energy as a power-law \cite{Caneva_PRB08} of the annealing time $\tau$. 
More precisely, one can define a residual energy per-spin, see Eq.~\eqref{eqn:resQC} and Sec.~\ref{ann:sec} for details,
and show \cite{Caneva_PRB08} that in the thermodynamics limit $N\to \infty$ one gets $\epsilon^{\res}(\tau) \sim \tau^{-3/2}$.
The situation changes drastically for $\prm\ge 3$ because the transition turns first-order \cite{Jorg_EPL10},
and the QA dynamics becomes very slow \cite{Bapst_JSTAT12}. 
Although methods have been proposed to avoid first-order transitions for this model using non-stoquastic Hamiltonians \cite{Seki2012,Seoane2012,NishimoriTakada2017,Susa2017}, we focus our attention to the traditional case of a stoquastic Hamiltonian with a transverse field.

The goal of the present paper is to precisely quantify how much QA is slowed down through an accurate determination of   
the thermodynamic-limit behaviour of the residual energy for $\prm\ge 3$.
Our conclusion, based on a finite-size analysis supplemented by the 
construction of the geometric envelope of the finite-$N$ Landau-Zener data, is that the thermodynamic limit of the
residual energy is logarithmic, $\epsilon^{\res}(\tau) \sim 1/\log{(\gamma \tau)}$ for any finite $\prm$. 
Interestingly, we show that an imaginary-time Schr\"odinger annealing dynamics displays, instead, a power-law behaviour 
$\epsilon^{\res}(\tau) \sim 1/\tau^2$ for all finite values of $\prm$, as predicted by the adiabatic theorem of quantum 
mechanics \cite{Morita_JMP08}. 

On the classical side, no study, to our knowledge, has tackled the single-spin-flip classical dynamics of the ferromagnetic
fully-connected $p$-spin model. This is the second important goal we set for our study. 
We solved this problem
by studying the deterministic evolution of a classical Master equation where the temperature $T$ is allowed
to change in time, in the spirit of Refs.~\onlinecite{Nishimori_PRE14,Zanca_2016} where the same approach is applied to the 
Ising model in one dimension. 
We find that the final asymptotic residual energy depends significantly on the final temperature $T_{\rmf}$ we set for the annealing.
For $\prm=2$, the classical Master equation annealing ends up being exponentially fast, $\epsilon^{\res}(\tau) \sim \nep^{-\tau/\tau^*}$,
hence winning over QA. For $\prm\ge 3$, however, the result is opposite: we have evidence that a classical annealing dynamics
{\em remains stuck}, for $N\to \infty$, in the wrong (paramagnetic) sector, never attaining, even for arbitrarily long
annealing times $\tau$, the correct minimal energy ferromagnetic state. 
%
%
Table \ref{table:summary} contains a summary of our results, together with some previously known facts about this problem.   

This paper is organized as follows. In Sec.~\ref{sec:model} we present the model we want to study and the different annealing
approaches we employed.
Section~\ref{sec:results} contains a detailed account of our results concerning the battle between Schr\"odinger QA and classical 
simulated annealing. 
Section~\ref{sec:results_IT} deals with the imaginary-time Schr\"odinger QA.
Section~\ref{sec:conclusions} contains a discussion and our conclusions. 
The Appendix contains a few technical details. 

\section{Annealing protocols} \label{sec:model}
%
Before embarking on the discussion of the results, we briefly introduce below the different types of annealing we will consider: 
a Schr\"odinger evolution quantum annealing, Sec.~\ref{qa:sec}, a thermal annealing of the classical Glauber-like 
master equation, Sec.~\ref{sa:sec}, and an imaginary-time Schr\"odinger quantum annealing, Sec.~\ref{itqa:sec}. 
Section~\ref{ann:sec} details how the annealing parameters are changed, and how the residual energy is defined in the different cases. 

\subsection{Quantum Annealing (QA-RT)}  \label{qa:sec}
To perform a QA dynamics on the fully-connected ferromagnetic $p$-spin model, we supplement the classical energy 
$H_{\rm C}$ in Eq.~\eqref{H_C:eqn} with the standard transverse field term as in Eq.~\eqref{eq:H_Q} which we
repeat here for the reader's convenience:
\begin{equation}\label{eq:H_Qbis}
{\widehat H}_{\rm Q}(t)=-\frac{JN}{2}\left(\frac{1}{N}\sum_{j=1}^N \PauliSigma^z_j \right)^{\prm} 
- \Gamma(t)\sum_{i=1}^N \PauliSigma _i^x\;.
\end{equation}
Here $\Gamma(t)$ is the time-dependent transverse field which we take to decrease towards $\Gamma_{\rmf}=0$ 
starting from some large value $\Gamma_{\rmi}$. 
Details on the annealing schedule will be given in Sec.~\ref{ann:sec}, see Eq.~\eqref{eqn:annQC}.
For very large $\Gamma_{\rmi}$, the initial ground state $|\psi_0\rangle$ is magnetized along the $\hat{x}$ spin direction and thus 
very disordered along the $\hat{z}$ direction.
The equilibrium properties of this model are well established, see for instance Refs.~\onlinecite{Bapst_JSTAT12,Jorg_EPL10}.
For what concerns the present work, it is important to recall that the model has a $2^{\rm nd}$ order
phase transition for $\prm=2$ at $\Gamma_c^{(\prm=2)}=J$, characterised by a critical-point gap scaling as 
$\Delta_{N}^{(\prm=2)} \sim N^{-1/3}$, and a $1^{\rm st}$ order phase transition for $\prm \geq 3$ with a gap which 
now scales exponentially with $N$, as $\Delta_{N}^{(\prm)} \sim \nep^{-\alpha_{\prm} N}$.
For increasing values of $\prm$ the critical transverse field $\Gamma_c^{(\prm)}$ approaches the Grover limit 
$\Gamma^{(\prm\to\infty)}_c=J/2$. 
The nature of the transitions, as a function of the temperature $T$, is identical in the classical case, as an elementary exact 
calculation shows:  $2^{\rm nd}$ order for $\prm=2$, which turns into $1^{\rm st}$ order for $\prm\ge 3$.
 
One quickly realises that since the Hamiltonian commutes with the total spin $\widehat{S}^2$
the study of the QA dynamics can be reduced to the Hilbert space subsector with the largest spin $S=N/2$, 
which is $(N+1)$-dimensional, see Ref.~\onlinecite{Caneva_PRB08,Bapst_JSTAT12}. 
Denoting the usual angular momentum states $|S=\frac{N}{2},M\rangle$, where 
$M=-\frac{N}{2},-\frac{N}{2}+1,\cdots,\frac{N}{2}$, with the shorthand $|m=\frac{2M}{N} \rangle$,
the amplitudes $\psi(m,t)=\langle m|\psi(t)\rangle$ of the time-evolving state vector $|\psi(t)\rangle$
obey a simple Schr\"odinger equation of the form (here and henceforth we set $\hbar=1$):
\begin{eqnarray} \label{eqn:Schroedinger-RT}
i \frac{\partial}{\partial t} \psi(m,t) &=& -\frac{N}{2} J m^{\prm} \psi(m,t) \nonumber \\
&& - \frac{N}{2} \Gamma(t) \sum_{\alpha = \pm 1} K_{m}^{(\alpha)} \, \psi(m-{\textstyle \frac{2\alpha}{N}},t) \;.
\end{eqnarray}
Here the first term originates from the classical (potential) energy $H_{\rm C}$, while the second (kinetic) term involves
$K_{m}^{(\pm)}=\sqrt{1-m^2 + 2(1 \mp m)/N}$, the square-root originating from the well-known angular momentum 
relationship $\opS^{\pm} |S,M\rangle= \sqrt{S(S+1)-M(M\pm 1)} | S,M\pm 1\rangle$ after rescaling by $2/N$.
%

\subsection{Simulated Annealing (SA)} \label{sa:sec}
To tackle the single-spin-flip classical dynamics of the ferromagnetic fully-connected $p$-spin model in Eq.~\eqref{H_C:eqn},
one needs to write down a classical Master equation (ME), in the spirit of the celebrated Glauber approach
to the one-dimensional Ising model \citep{Glauber_JMP63}, for the probability $P(\sigma,t)$ that the system is in 
configuration $\sigma$ at time $t$.
In continuous time, such a ME would have the following linear form\cite{vanKampen:book}:    
\begin{equation} \label{eqn:ME}
\frac{\partial P(\sigma,t)}{\partial t} =
\sum_{\sigma'} W_{\sigma,\sigma'} P(\sigma',t) - \sum_{\sigma'} W_{\sigma', \sigma} P(\sigma,t) \;.
\end{equation}
Here $W_{\sigma,\sigma'}$ is the ``rate'' matrix describing the transition from a configuration $\sigma'$ to $\sigma$,
and the second term describes that inverse process $\sigma\to \sigma'$. 
The configurations $\sigma'$ which are connected to $\sigma$ by $W_{\sigma,\sigma'}$ can be regarded
as {\em neighbours} of $\sigma$: we will consider only {\em single spin-flip} moves, denoting with
$\sigma'=\sigmabar^j=(\sigma_1,\cdots,-\sigma_j,\cdots,\sigma_N)$ the configuration which differs from $\sigma$
by a flip of the variable $\sigma_j$. 
Even restricting in this way the ``neighbours'' $\sigma'$, there is still a large freedom in the choice of $W_{\sigma,\sigma'}$. 
Typically, one imposes that the choice is such that the ME converges, when the temperature is kept fixed, 
towards the equilibrium Boltzmann probability distribution $P^{\rm eq}(\sigma)$. 
An effective way to impose this approach to equilibrium is to require that the {\em detailed balance condition} (DB) is satisfied: 
\begin{equation} \label{eqn:DB}
W_{\sigma,\sigma'} P^{\rm eq}(\sigma') = W_{\sigma',\sigma} P^{\rm eq}(\sigma) \;.
\end{equation}
Even imposing the DB condition leaves a considerable freedom in the choice of $W_{\sigma,\sigma'}$.
One possible choice is the usual Metropolis rule, very common in Monte Carlo studies, which we will not
consider here because it is not analytically very convenient.
A second widely used form of $W_{\sigma,\sigma'}$ satisfying DB, which we will adopt in the following, 
is the {\em heat bath} choice: 
\begin{eqnarray} \label{eqn:Whb}
W_{\sigma,\sigma'} &=&
\alpha_0 \frac{\nep^{-\beta H_{\rm C}(\sigma)} }{\nep^{-\beta H_{\rm C}(\sigma)} + \nep^{-\beta H_{\rm C}(\sigma')} } \nonumber \\
&=& \alpha_0 \frac{\nep^{-\frac{\beta}{2}\Delta E_{\sigma,\sigma'}}}{\nep^{-\frac{\beta}{2}\Delta E_{\sigma,\sigma'}}+\nep^{\frac{\beta}{2}\Delta E_{\sigma,\sigma'}}} \;,
\end{eqnarray}   
where $\Delta E_{\sigma,\sigma'}=H_{\rm C}({\sigma})-H_{\rm C}({\sigma'})$ is the classical energy difference in
changing the configuration from $\sigma'$ to $\sigma$, and $\alpha_0$ is an overall rate constant which we easily reabsorb
in our units of time.

Following the approach of Ref.~\onlinecite{Nishimori_PRE14}, as recently applied on the random Ising chain problem
in Ref.~\onlinecite{Zanca_2016}, we might transform a classical master equation into an equivalent {\em imaginary-time}
Schr\"odinger problem with an appropriate effective Hamiltonian ${\mathcal H}_{\sigma,\sigma'}$ which effectively
``symmetrizes'' the rate matrix $W_{\sigma,\sigma'}$ using DB; from there, one would then proceed to study such 
equivalent imaginary-time Schr\"odinger problem using the same ``total spin technique'' employed above for the quantum case. 
One might do that, but we will not do it here, for a reason that we briefly explain in Appendix \ref{app:mapping}.

Now, we will directly exploit the ``permutation symmetry'' of the classical problem to do something that is completely 
equivalent to working in the ``maximum spin sector''.
Indeed, a common feature of the many possible choices of $W_{\sigma,\sigma'}$ based on DB is that $W_{\sigma,\sigma'}$ 
depends on the configurations $\sigma$ and $\sigma'$ only through their classical energies $H_{\rm C}(\sigma)$
and $H_{\rm C}(\sigma')$.  
These, in turn, for the model we are considering, depend only on the total magnetization $m=\frac{1}{N}\sum_j \sigma_j$:
$H_{\rm C}(\sigma)= -JN m^{\prm}/2 \equiv E(m)$, where we introduce the shorthand $E(m)$ to denote the classical energy
of a configuration with magnetization (per spin) $m$.
Hence, we can regard $W_{\sigma,\sigma'}={\mathbb W}_{m,m'}$ and make the {\em Ansatz} that $P(\sigma,t)$ itself 
will depend on $\sigma$ only through $m$, provided we account for the appropriate combinatorial factors. 
This transforms the classical ME into a workable problem of the same difficulty as the Schr\"odinger equation in 
Eq.~\eqref{eqn:Schroedinger-RT}.
If $N_+$ is the number of $\uparrow$ spins in the configuration $\sigma$, and $N_-$ that of $\downarrow$ spins,
we can express the magnetization $m$ as $m=(2N_+-N)/N\in[-1,1]$. 
For finite $N$, $m$ can assume only $N+1$ values which differ by $2/N$, i.e., $m=-1+\frac{2k}{N}$ with $k=0\cdots N$.
The number of energetically and dynamically equivalent configurations $\sigma$ corresponding to a magnetization
$m$ being given by $\binom{N}{N_+}$, we can define the probability distribution for the magnetization $m$ as:
\begin{equation}
\Pm(m, t) = \binom{N}{N_+} P(\sigma,t) \;.
\end{equation}
Rewriting the classical ME Eq.~\eqref{eqn:ME} in terms of $\Pm(m,t)$, with due account of all the binomial factors, 
we eventually arrive at the following ``permutation symmetric'' version of it:
\begin{eqnarray} \label{eqn:MEsymm}
-\frac{\partial \Pm(m,t)}{\partial t} &=& -\frac{N}{2} \sum_{\alpha=\pm} {\mathcal W}_m^{(\alpha)} \, \Pm(m-{\textstyle \frac{2\alpha}{N}},t) 
\nonumber \\ 
&& + \frac{N}{2} \, {\mathcal V}_{m} \, \Pm(m,t) \;.
\end{eqnarray}
Again, $\frac{2}{N}$ is the change in magnetisation upon flipping a single spin, but the effective
kinetic and potential term coefficients are now given by: 
\begin{equation}
\left\{
\begin{array}{lcl}
{\mathcal W}_m^{(\alpha)} &=& (1+\alpha m+{\textstyle \frac{2}{N}}) \, {\mathbb W}_{m,m+\frac{2\alpha}{N}} \vspace{3mm} \\
{\mathcal V}_m &=& \displaystyle \sum_{\alpha =\pm}  (1+\alpha m) \, {\mathbb W}_{m-\alpha\frac{2}{N},m}\
\end{array} \right. \;.
\end{equation}
%
Notice that the transition rates ${\mathbb W}_{m,m'}$ depend on the energy difference $E(m)-E(m')$ and also on 
the inverse temperature $\beta=1/(k_BT)$. 
To do a SA dynamics we have to ``anneal down'' the temperature $T(t)$ entering in the heat-bath transition rates, thus
making all the relevant ingredients entering Eq.~\eqref{eqn:MEsymm} {\em time-dependent}:
${\mathbb W}_{m,m'}(T(t))$,  ${\mathcal W}_m^{\alpha}(T(t))$, and ${\mathcal V}_m(T(t))$. 
Eq.~\eqref{eqn:MEsymm}, with the prescribed time-dependent coefficients ensuing from the time-dependence of $T(t)$,
is the linear system of $N+1$ coupled differential equations that we will need to solve to do SA dynamics for the fully-connected
spin ferromagnet, in the spirit of a Glauber-type classical master equation.

\subsection{Schr\"odinger quantum annealing in imaginary time (QA-IT)} \label{itqa:sec}

The analogy of the classical ME in Eq.~\eqref{eqn:ME} with an imaginary-time Schr\"odinger problem is well known\cite{vanKampen:book}:
it can be made precise by a suitable symmetrization of the transition rate matrix $W_{\sigma,\sigma'}$ to transform it into a proper Hermitean 
quantum kinetic energy. 
Such an analogy is inspiring, as it correctly suggests, for instance, that the thermal annealing dynamics proceeds 
by ``filtering-out'' the higher excited eigenstates from the time-evolving $P(\sigma,t)$.
It should, however, not be confused with the actual {\em imaginary-time} (IT) Schr\"odinger dynamics obtained by a Wick's rotation 
$t\to -it$ of Eq.~\eqref{eqn:Schroedinger-RT}, which amounts to studying:
\begin{eqnarray} \label{eqn:Schroedinger-IT}
-\frac{\partial}{\partial t} \psi(m,t) &=& -\frac{N}{2} J m^{\prm} \psi(m,t) \nonumber \\
&& - \frac{N}{2} \Gamma(t) \sum_{\alpha = \pm 1} K_{m}^{(\alpha)} \psi(m-{\textstyle \frac{2\alpha}{N}},t) \,. \hspace{4mm}
\end{eqnarray}
Although physically not relevant for the actual hardware of possible QA machines, this route is interesting from the algorithmic
point of view, as many quantum Monte Carlo approaches are indeed based on an IT framework. 
Moreover, the filtering effect towards the ground state of the IT dynamics is beneficial within an optimization context.
Indeed, recent results for an Ising chain \citep{Zanca_2016} suggest that the residual energies 
obtained using QA in IT are definitely below those of a standard QA in {\em real time} (RT).   

\subsection{The annealing schedule and the residual energy} \label{ann:sec}
To fully specify the annealing, we have to stipulate the annealing schedule we will use for the relevant parameters:
the transverse field $\Gamma(t)$ for QA, and the temperature $T(t)$ for SA.
Although a schedule optimization is known to be highly important \cite{Morita_JMP08,Herr2017} --- for instance, it 
provides the quadratic quantum speedup in a QA version of the Grover's search problem \cite{Roland_PRA02} --- 
for the purpose of a simpler setting we will adopt here a linear schedule for both QA and SA, writing:
\begin{equation} \label{eqn:annQC}
\left\{ \begin{array}{ll}
\displaystyle \Gamma(t) =\Gamma_{\rmi} \left(1-\frac{t}{\tau}\right) + \Gamma_{\rmf} \frac{t}{\tau}  & \hspace{4mm} {\rm QA} \vspace{3mm} \\
\displaystyle T(t) =T_{\rmi}\left(1-\frac{t}{\tau}\right) + T_{\rmf} \frac{t}{\tau} & \hspace{4mm} {\rm SA}
\end{array} \right.
\end{equation}
On equal footing with QA, where the initial state is the ground state of a suitably large $\Gamma_{\rmi}>\Gamma_c$, 
the SA evolution will start from the equilibrium configuration at a sufficiently large $T_{\rmi}\gg T_c$, where $T_c$ is
the equilibrium critical temperature.  
Notice that we will allow ourselves a bit more freedom in the SA case, by choosing $T_{\rmf}$ to be the final temperature, 
which we could take to be either much less than $T_c$, or simply set to $T_{\rmf}=0$. 
We will see the role of this choice in discussing the SA results.  
In the QA case the value of $\Gamma_{\rmf}$ is much less relevant, as long as $\Gamma_{\rmf}\ll \Gamma_c$, and 
we simply set $\Gamma_{\rmf}=0$. 

To characterise the annealing efficiency we study, as often done, the residual energy (per spin) at the end of annealing,
defined as the difference between the total energy in the final state, minus the corresponding minimal energy.
In the quantum case, the definition involves the average of the final Hamiltonian $\widehat{H}_{\rm Q}(\tau)=H_{\rm C}$
over the time-evolved final state $ |\psi(\tau) \rangle$. 
In the classical case the corresponding quantity to calculate is the average of the classical $H_{\rm C}$ over the final 
probability distribution $P(\sigma,\tau)$. 
In both cases we write:
\begin{equation} \label{eqn:resQC}
\epsilon^{\res}_N(\tau) = \left\{ \begin{array}{ll}
\displaystyle \frac{1}{N} \left[ \frac{\langle \psi(\tau) | \widehat{H}_{\rm Q}(\tau) |\psi(\tau) \rangle}{\langle \psi(\tau) | \psi(\tau) \rangle} - E_0 \right]  
& \hspace{4mm} {\rm QA} \vspace{3mm} \\
\displaystyle \frac{1}{N} \Big( \sum_m E(m) \Pm(m,\tau) - \langle E \rangle_{\rm eq}  \Big)  
& \hspace{4mm} {\rm SA}
\end{array} \right.
\end{equation}
where $E_0$ is the final minimal energy value, $E_0=-NJ/2$ in the present case,
while $\langle E \rangle_{\rm eq}$ denotes the average energy over the equilibrium distribution at the final
temperature $T_{\rmf}$.
$\epsilon^{\res}_N(\tau)$ is in general a function of $\tau$ and of the number of sites $N$. 
Its dependence on $\prm$ is implicit, but, as we shall see, the value of $\prm$ plays a crucial role in the following.

\section{Results: QA-RT versus SA} \label{sec:results}
%
We present here the results of our analysis of both QA-RT and SA for the fully-connected $p$-spin Ising ferromagnet. 
We distinguish the $\prm=2$ case, where the transition (both classical and quantum) is second-order, 
from $\prm\ge 3$ where the classical and quantum transitions are first-order. 
(We will mostly present results for $\prm=3$, but the higher $\prm$ we have explored present a similar phenomenology.)
We will in the end discuss separately the Grover limit $\prm({\rm odd})=\infty$. 
In our numerical analysis we simulated the QA-RT and SA dynamics by annealing down the driving parameter 
with a constant rate in a total annealing time $\tau$, see Eq.~\eqref{eqn:annQC}.
To characterize the behaviour of the residual energy density Eq.~\eqref{eqn:resQC}, 
we solved the Schr\"odinger equation~\eqref{eqn:Schroedinger-RT}, and the classical ME Eq.~\eqref{eqn:MEsymm}, 
for several values of the total annealing time $\tau$ and increasing the number of sites $N$ up to 1024 spins.
A comparison between real-time and imaginary-time Schr\"odinger QA is deferred to Sec.~\ref{sec:results_IT}.
We set the coupling $J=1$ in what follows.

The QA-RT dynamics of the present model was previously studied in Ref.~\onlinecite{Caneva_PRB08} for $\prm=2$, 
finding $\epsilon^{\res}_{N\to\infty}(\tau) \sim \tau^{-3/2}$ in the thermodynamic limit,  
and in Ref.~\onlinecite{Bapst_JSTAT12} for general $\prm$, with an exhaustive analysis which however did not go all
the way to predicting the final asymptotic large-$\tau$ behaviour of $\epsilon^{\res}_{N\to\infty}(\tau)$.

\subsection{$\prm=2$} \label{sec:p2}
%
We start by presenting the results obtained for $\prm=2$. This case was already discussed in Ref.~\onlinecite{Caneva_PRB08}. The
analysis  performed in the present work will prove crucial in dealing with $\prm\ge 3$.
%
\begin{figure*}[t!]
\begin{center}
\includegraphics[width=175mm]{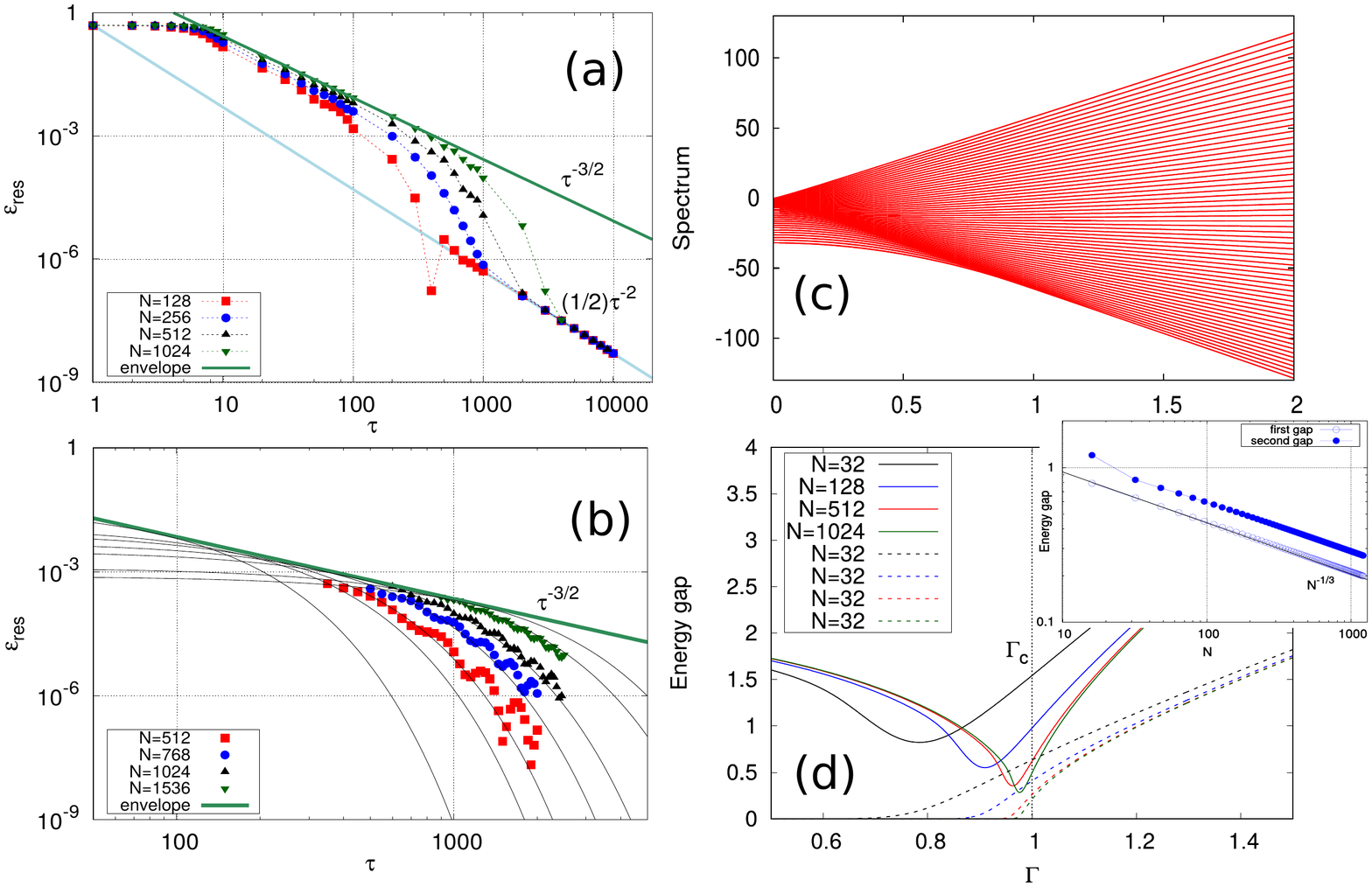}
\caption{(a) Log-log plot of the residual energy density {\it vs} annealing time for QA-RT for $\prm=2$. 
The QA-RT raw data show the four different regimes discussed in the text. 
Here $\Gamma_{\rmi} = 2$.
(b) The envelope construction illustrated: notice the large coherent oscillations of the finite-size LZ data, discussed in the text.
The green thick line represents the geometrical envelope of our QA-RT finite-$N$ curves, obtained from LZ fit (black solid lines).
(c) The spectrum for $N=64$. 
(d) The minimum gap $\Delta_N$ (dashed lines) and the more relevant dynamical gap (solid lines) {\it vs} the transverse field $\Gamma$. The inset shows the scaling with $N$ of the minimum gap $\Delta_N$ at the critical point. 
}
\label{fig:res_P2}
\end{center}
\end{figure*}
%
Fig.~\ref{fig:res_P2} shows our QA results for $\prm=2$ for Schr\"odinger annealing in real time (RT). 
As first reported by Caneva {\em et al.}~\cite{Caneva_PRB08}, the QA-RT data for $\epsilon^{\res}_N(\tau)$ as a function of the 
annealing time $\tau$ show four regimes: 
\begin{description}
\item[1a)] an initial very short-$\tau$ behaviour, where the system is essentially unable to follow the drive, and remain trapped in the
paramagnetic phase; 
\item[1b)] a subsequent power-law scaling which ``feels'' the critical point of the system, because the driving is too fast 
for the system to realize that, at finite-$N$, the gap $\Delta_N$ is non-vanishing, $\Delta_N>0$: 
here $\epsilon^{\rm res}(\tau)\sim \tau^{-3/2}$.
From Fig.~\ref{fig:res_P2}(c) and Fig.~\ref{fig:res_P3}(c), we see that the ground state for large $\Gamma$ is driven, upon decreasing $\Gamma$, 
through a series of avoided level crossings. 
\item[2)] an intermediate Landau-Zener regime where the driving is slow enough that the system ``sees'' that $\Delta_N>0$,
but not so slow to be completely adiabatic. 
Here the evolution is dominated by {\em single} Landau-Zener (LZ) events for an effective two-level-system describing the 
avoided crossing, with a gap $\Delta_N>0$, between the two lowest-lying instantaneous eigenvalues. 
Notice that for even values of $p$ the relevant gap is between the ground state and the {\em second} excited state, 
because $\hat{H}_{\rm Q}$ preserves parity.
The dynamical gap is therefore finite for $\Gamma<\Gamma_c$: it shares the same critical properties of the first equilibrium energy gap,
as shown in Fig.~\ref{fig:res_P2}(d).
The LZ formula predicts that the probability of being excited across such an avoided level crossing transition
goes like $P_{\rm LZ}=\nep^{-\frac{\pi}{4}\Delta_N^2\tau}$, which suggests that the residual energy per-spin in this 
intermediate regime should behave as:
\begin{equation} \label{eqn:epsresLZ}
\epsilon^{\LZ}_N(\tau)=\frac{C}{N}\nep^{-\tau/\tau^*_N} \;,
\end{equation}
where $\tau^*_N\propto \Delta_N^{-2}$ is a characteristic time for the LZ transition.
\item[3)] a final asymptotic regime for $\tau \gg \Delta_N^{-2}$ --- hence, only observed for finite $N$ --- 
which satisfies the adiabatic theorem prediction~\cite{Suzuki_JPSJ05,Morita_JMP08} of $\epsilon^{\rm res}(\tau)\propto 1/\tau^2$. 
In the present case we find, for general value of $\prm$:
\begin{equation} \label{eqn:eres_adiabatic}
\epsilon^{\rm res}(\tau\gg \Delta_N^{-2}) \approx \frac{\Gamma_{\rmi}^2}{p^3} \frac{1}{\tau^2} \;. 
\end{equation}
\end{description}

Although the final asymptotic behaviour of $\epsilon^{\res}$ for $N\to\infty$ turns out to be, in the end, given by the 
power-law scaling $\tau^{-3/2}$ discussed in {\bf 1b)}, the most interesting, and in some sense revealing, 
regime is the intermediate-$\tau$ LZ region {\bf 2)}. 
This LZ-regime will be present also for $\prm\ge 3$ and shows a behaviour of the residual energy which can be cast in the form:
\begin{equation} \label{eqn:epsresLZ_full}
\epsilon^{\LZ}_N(\tau)= \left\{
\begin{array}{lcl} 
\displaystyle \frac{C_2}{N} \nep^{-\gamma_2 \tau \,N^{-2z} } & \mbox{for} & \prm=2 \vspace{3mm}\\
\displaystyle \frac{C_\prm}{N}\nep^{-\gamma_{\prm} \tau \, \nep^{-2\alpha_\prm N} } & \mbox{for} & \prm\ge 3
\end{array} \right.\;.
\end{equation}
Here $C_{\prm}$ is, as we have verified, very close to the first energy gap at the end of the annealing, 
i.e, $C_{\prm}\approx \prm J + O(1/N)$. 
The difference between $\prm=2$ and $\prm\ge 3$ resides in the way the transition gap
$\Delta_N$ closes with increasing $N$: 
as a power-law~\cite{Caneva_PRB08} $\Delta_N\sim N^{-z}$ (with $z=1/3$) for $\prm=2$ ($2^{\rm nd}$-order transition), 
or exponentially~\cite{Jorg_EPL10,Bapst_JSTAT12} $\Delta_N\sim \nep^{-\alpha_p N}$ for $\prm\ge 3$ ($1^{\rm st}$-order transition).  
To obtain the thermodynamic limit asymptotic behaviour of $\epsilon^{\res}_{N\rightarrow \infty}(\tau)$ 
we resort to a strategy that will prove extremely useful when $\prm\ge 3$ 
--- less so for $\prm=2$, where the asymptotic is already announced by the intermediate power-law regime $\tau^{-3/2}$ --- 
but we choose to illustrate here.

The strategy is based on the geometrical construction of the {\em envelope} of all the finite-$N$ LZ-curves, 
and is briefly described in Appendix \ref{app:envelope} where we illustrated it with the example of the transverse-field Ising chain.
Briefly, if $f_{u}(x)$ is a family of functions which smoothly depends on some parameter $u$, 
its envelope $e(x)$ is a function which is tangent in each point to a member of the family, a condition that is enforced
by solving $\partial_u f_u(x) = 0$ to find $u(x)$ and then setting $e(x)=f_{u(x)}(x)$.
In our case, we identify $x\mapsto \tau$, $u\mapsto N$ (which we assume to be real, rather than integer), 
and $f_u(x)\mapsto \epsilon^{\LZ}_N(\tau)$. 
To construct the envelope $\epsilon^{\rm env}(\tau)=\epsilon^{\LZ}_{N(\tau)}(\tau)$ we need to solve for $N(\tau)$ the 
implicit equation $\partial_N \epsilon^{\LZ}_N(\tau)=0$. For $\prm=2$ the solution is very simple:
\begin{eqnarray} \label{eqn:env_p2}
N(\tau) &=& (2z\gamma_2 \tau)^{\frac{1}{2z}} \\
\epsilon^{\rm env}(\tau) &=& \frac{C_2}{N(\tau)} \nep^{-2z} = 
\left(\frac{3}{2\,\nep\gamma_2}\right)^{\frac{3}{2}} \frac{C_2}{\tau^{\frac{3}{2}}}\;,			
\end{eqnarray}
where we have used the fact that $z=1/3$. 
As shown in Fig.~\ref{fig:res_P2}(b), the method works well, although here 
--- at variance with the transverse-field Ising chain illustrated in Appendix \ref{app:envelope} --- 
the LZ-regime is ``decorated'' by extra coherent oscillations which make the analysis of small-$N$ data a bit more difficult:
notice that these oscillations become less and less pronounced when $N$ increases.   
The construction allows us to predict quite precisely both the scaling exponent $3/2$ and even the numerical coefficient in front of 
the power-law from the sole analysis of the LZ-regime. 

\begin{figure*}[ht!]
\begin{center}
\includegraphics[width=180mm]{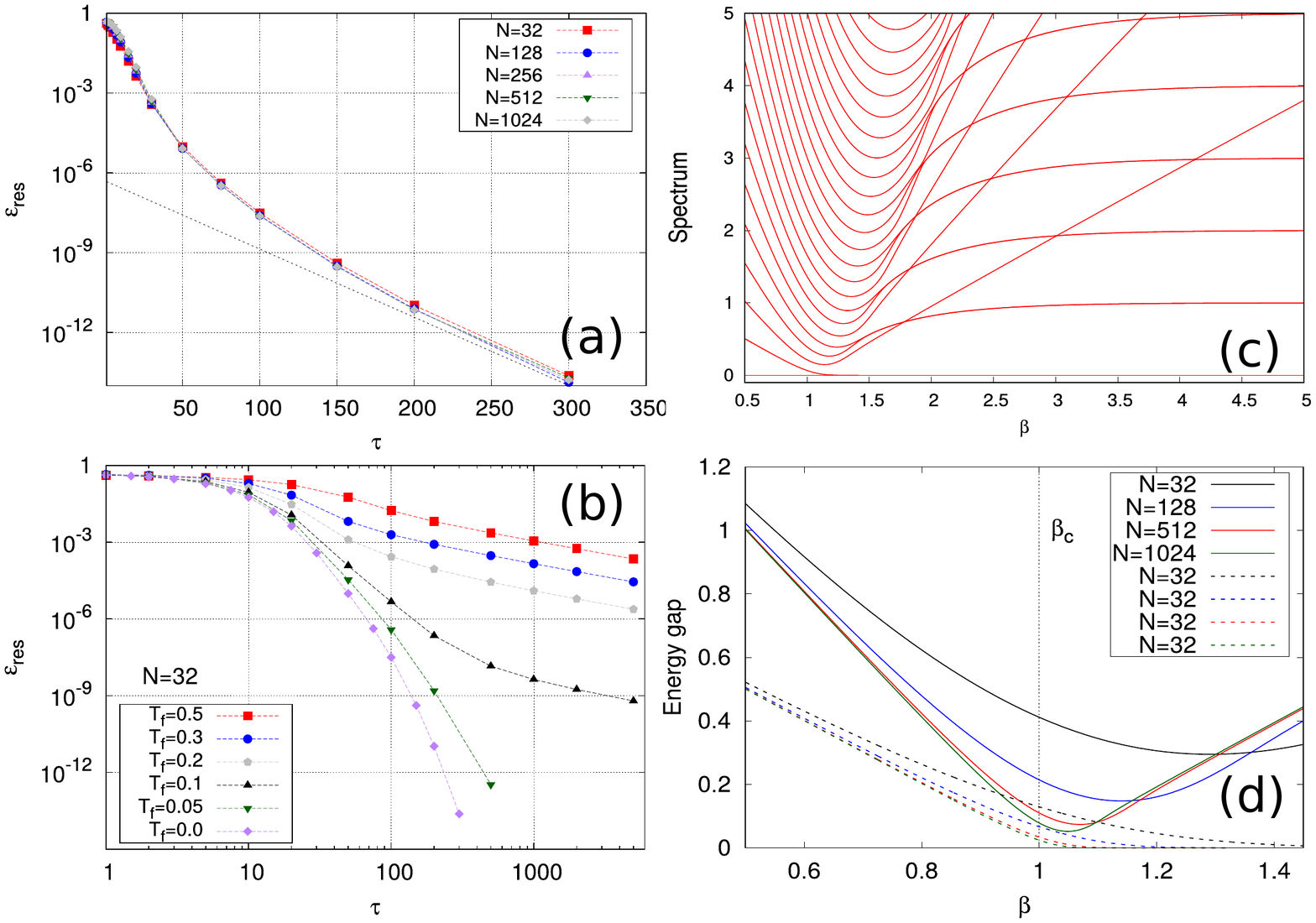}
\caption{
(a) $\epsilon^{\res}_N(\tau)$ vs annealing time $\tau$ for SA at $\prm=2$, with vanishing final temperature $T_{\rmf}=0$. 
The data show almost no size dependence and follow asymptotically an exponential relaxation. 
(b) $\epsilon^{\res}_N(\tau)$ for $N=32$ in log-log scale, for different values of the final temperature $T_{\rmf}$. 
In (a) and (b) the initial temperature is $T_{\rmi}=2$. 
For $T_{\rmf}>0$ the asymptotic adiabatic regime $\tau^{-1}$ suggested by Eq.~\eqref{eq:ad_CA} is visible.
(c) The instantaneous spectrum of the classical master equation for $N=128$. 
(d) The minimum gap relevant for the classical master equation dynamics (solid line) and for equilibrium properties (dashed lines).
}
\label{fig:res_P2SA}
\end{center}
\end{figure*}
%
Let us now illustrate our results for the SA dynamics. 
First of all, the residual energy depends crucially on the final annealing temperature
$T_{\rmf}$, leading to very different results depending on whether $T_{\rmf}>0$ or $T_{\rmf}=0$.
Results for $T_{\rmf}=0$ and $T_{\rmf}>0$ are reported in Figs.~\ref{fig:res_P2SA}(a) and \ref{fig:res_P2SA}(b), respectively. 
Notice that the data bear no memory of the properties of the critical point, as the dynamical spectral gap between the 
lowest eigenstate and the first accessible excited state opens-up again after the critical point crossing, 
see spectrum and spectral gaps in Fig.~\ref{fig:res_P2SA}(c,d).
The ``filtering'' effect is therefore very effective and we observe a size-independent decrease of $\epsilon^{\res}$ with $\tau$, 
which is compatible with an exponential decay for large annealing times $\tau$. 
Notice also the striking absence of any asymptotic adiabatic regime, as opposed to the $1/\tau^2$ behaviour of QA.
Indeed, had we stopped our SA at a final $T_{\rm f}>0$ we would have observed an adiabatic power-law tail of the form
\begin{eqnarray} \label{eq:ad_CA}
\epsilon^{\res}(t=\tau) &\simeq& \langle \phi_0 | H_{\rm C} | \phi_{\rm ex}\rangle \; c_{\rm ex}(\tau) \nonumber \\
c_{\rm ex}(\tau) &=& \frac{2T_{\rmi}}{\tau} 
\frac{\langle \phi_{\rm ex}| \partial_{\beta} \calH |\phi_0 \rangle\big|_{T=T_{\rm f}}}{\Delta^2_{\rm ex}} \;,
\end{eqnarray}
where $c_{\rm ex}(\tau)$ is the projection of the probability $\Pm(m,t=\tau)$ over the first relevant lowest-lying 
left eigenstate of the transition matrix $W_{\sigma,\sigma'}$  (see Appendix~\ref{app:adiab} for details).
Differently from the usual result of the adiabatic theorem \cite{Morita_JMP08}, the power-law pre-factor 
$\langle \phi_{\rm ex}| \partial_{\beta} \calH |\phi_0 \rangle$ decreases exponentially fast as $T_{\rmf}\to 0$, see Eqs.~\eqref{eq:Heff_hb},\eqref{eq:Heff_hb_der}.
As it turns out, the ultimate behaviour of SA for $T_{\rmf}=0$ is exponential.
What makes SA so different from QA is the dependence of the Hamiltonian from the annealing parameter:
in QA $\widehat{H}_{\rm Q}$ depends linearly on the transverse field $\Gamma$, while
in SA the dependence of the transition rates ${\mathbb W}_{m,m'}$ 
from the temperature $T$ is highly non-linear, through the Boltzmann weights.

\subsection{$\prm\ge 3$} \label{sec:p3}
\begin{figure*}[ht!]
\begin{center}
\includegraphics[width=180mm]{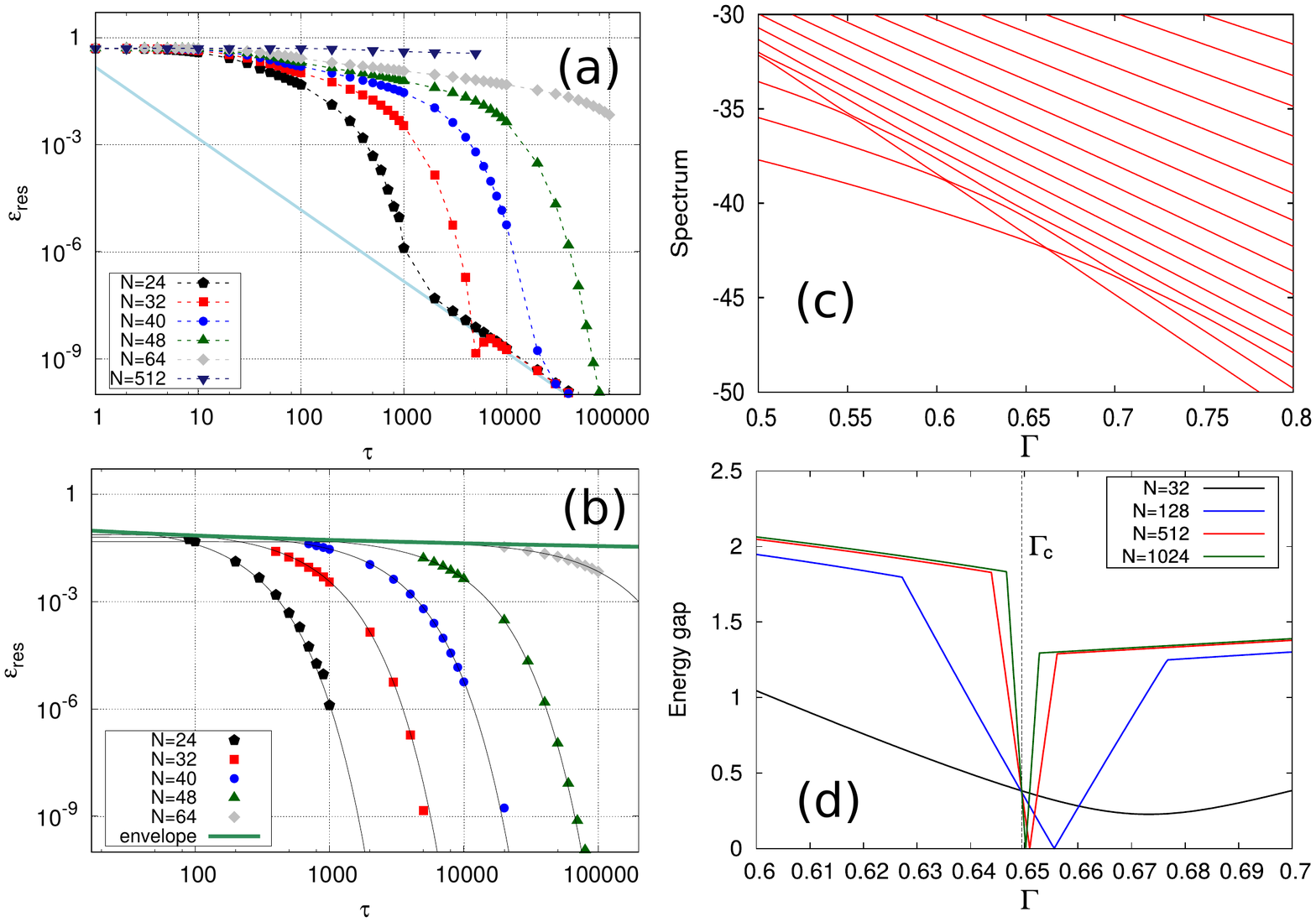}
\caption{(a) Log-log plot of the residual energy density vs annealing time for QA-RT for $\prm=3$. 
For first order phase transition the characteristic annealing times $\tau^*_N$ increases exponentially with $N$,
hence only for moderate $N$ the LZ-regime is clearly visible. Here the initial transverse field is set to $\Gamma_{\rmi}=2$.
(b) Detail of the exponential fit on the regions in which the LZ approximation holds, used for the envelope construction. 
The envelope $1/\log(\gamma	\tau)$ (thick line) gives an estimate of the large-$\tau$ behaviour of the residual energy in the thermodynamic limit $N\to \infty$.
(c) The instantaneous spectrum vs the transverse field $\Gamma$, close to the transition, for $N=64$.
One can clearly see the series of avoided level crossings encountered by the paramagnetic phase coming from the right.
(d) The minimum gap close to the critical point.}
\label{fig:res_P3}
\end{center}
\end{figure*}

%
In this case the corresponding equilibrium model undergoes a first-order phase transition, as compared to the case 
$p=2$ where the transition is of second order. This difference has an enormous impact on the efficiency of the annealing protocols.
Let us now focus on the most interesting situation $\prm\ge 3$, where the transition is first order and 
the optimization process becomes hard \cite{Jorg_EPL10,Bapst_JSTAT12}.
Here we observe three distinct regimes for the QA-RT evolution, see Fig.~\ref{fig:res_P3}(a). 
The first (short $\tau$) and the third (asymptotic adiabatic behaviour) are strict analogues of regime {\bf 1a)} and {\bf 3)} 
observed for $\prm=2$.
The intermediate LZ-regime {\bf 2)} is now very clear, but only visible in a limited range of $N\sim 24\div 64$: for larger $N$ it
would occur in a region of annealing times $\tau$ prohibitively hard to simulate, since the critical gap $\Delta_N$, 
and therefore the characteristic time $\tau^*_N\sim \Delta_N^{-2}$, now scale exponentially\cite{Jorg_EPL10,Bapst_JSTAT12} 
with $N$, see Eq.~\eqref{eqn:epsresLZ_full}.
The absence of coherent oscillations observed in Sec.~\ref{sec:p2} is due to the sharp closing of the gaps between successive eigenvalues of the spectrum
 ---Figs.~\ref{fig:res_P3}(c) and (d)--- which rapidly decouples the two states involved in the LZ approximation.
The analytical construction of the envelope is now a bit more involved, since the equation defining $N(\tau)$ --- the implicit solution
of $\partial_N \epsilon^{\LZ}_N(\tau)=0$ --- cannot be solved explicitly. Nevertheless, one can show that:
\begin{eqnarray} \label{eqn:eres_qa_rt_p3}
\frac{\nep^{2\alpha_{\prm} N(\tau)}}{2\alpha_{\prm} N(\tau)} &=& \gamma_{\prm} \tau \nonumber \\
\epsilon^{\rm env}(\tau) &\approx& \frac{2\alpha_{\prm} C_{\prm}}{\log(\gamma_{\prm}\tau)+\log(\log(\gamma_{\prm}\tau))}
\nep^{-\frac{1}{\log(\gamma_{\prm}\tau)}}\,, \hspace{2mm}
\end{eqnarray}
where, see Eq.~\eqref{eqn:epsresLZ_full}, $\alpha_{\prm}$ enters in the exponential closing of the critical gap 
$\Delta_N\sim \nep^{-\alpha_{\prm} N}$ and $\gamma_{\prm}$ is the overall rate-constant entering the LZ 
expression~\eqref{eqn:epsresLZ_full}.
($C_{\prm} \simeq \prm J$ is again very close to the lowest gap $\Delta(\Gamma=0)$ at the end of the 
annealing process.)

One should pause here to appreciate the power of the envelope construction. 
If we look at the raw data in Fig.~\ref{fig:res_P3}(a), it would be hopeless to try to {\em fit} the slight (but evident) decline of the 
$N=512$ annealing data with some inverse power of a logarithm of $\tau$ plus (unknown) corrections. 
But if you extract (by a simple fitting) the relevant ingredients from the LZ-regime of the annealing data for 
moderate $N = 24\div 64$, see Fig.~\ref{fig:res_P3}(b), this will unambigously signal, if you assume that an envelope exists, 
the asymptotic result in Eq.~\eqref{eqn:eres_qa_rt_p3}.
The envelope construction is, in a way, a powerful ``telescope'' for data that you would never be able to observe directly.   
Summarising, consistently with Refs.~\onlinecite{Jorg_EPL10,Bapst_JSTAT12}, first-order transitions lead to exponentially small
gaps and to a problem that is hard for QA, but we can precisely quantify this hardness by stating that, for $N\to \infty$, 
the residual error decreases as $\epsilon^{\res}(\tau) \sim 1/\log(\gamma_{\prm} \tau)$ as the annealing time $\tau$ grows.
Such a statement applies to all {\em finite} values of $\prm$. The Grover limit should be considered separately. 
%

\begin{figure*}[ht!]
\begin{center}
\includegraphics[width=180mm]{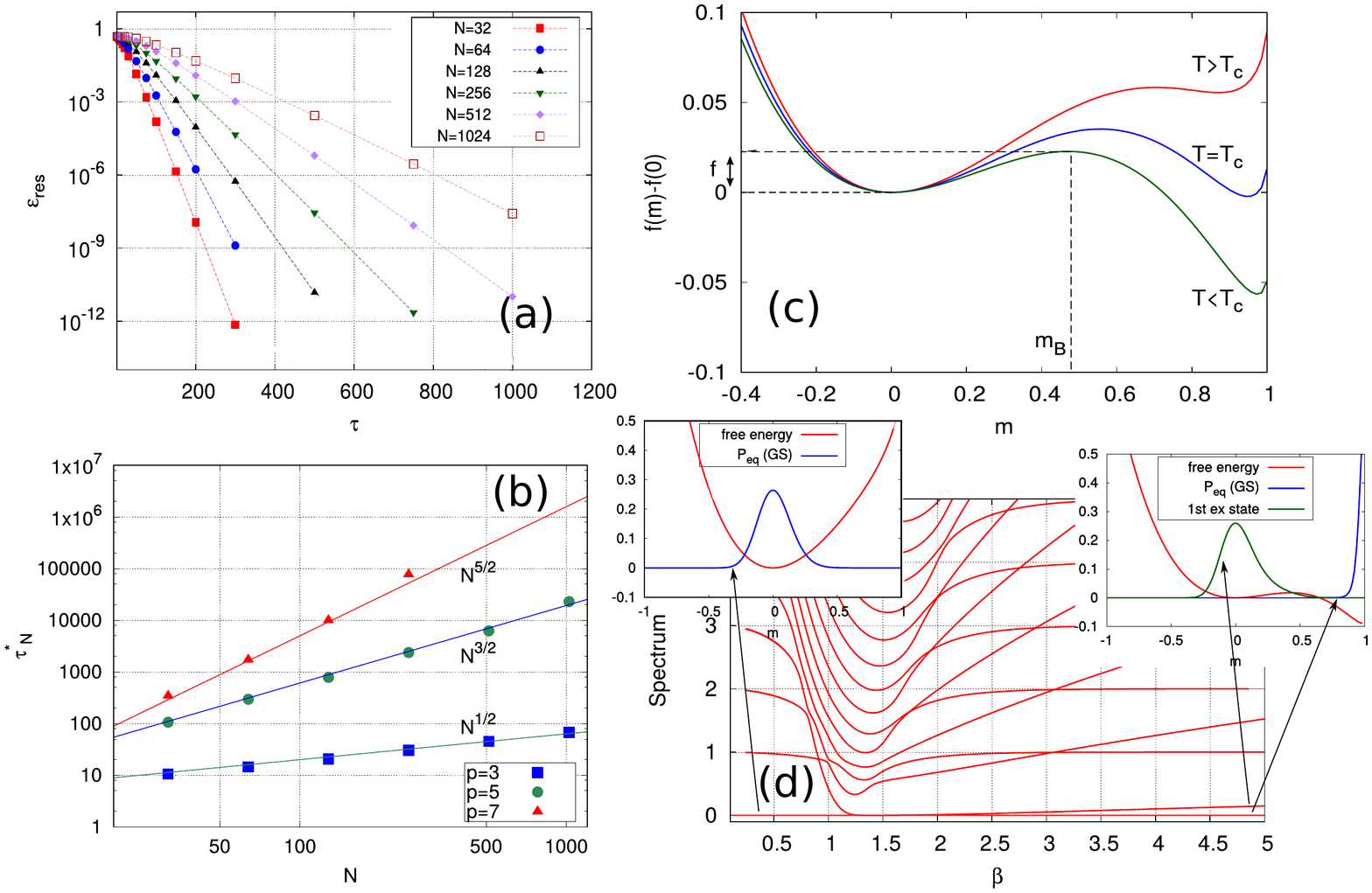}
\end{center}
\caption{(a) $\epsilon^{\res}_N(\tau)$ for SA with $T_{\rmi}=2$, $T_{\rmf}=0$ when $\prm=3$.
Differently from the case $\prm=2$, the relaxation rate $1/\tau^*_N$ depends on $N$ and vanishes with a power law 
in the thermodynamic limit. 
(b) Scaling of $\tau^*_N$ with $N$ for vanishing final temperature $T_{\rmf}$, compared with the power laws obtained analytically.
(c) The classical equilibrium free-energy density  $f(m,T)$ for the $\prm=3$ fully-connected $p$-spin model
vs the magnetization $m=(\sum_j \sigma_j)/N$ for different temperatures.
(d) The instantaneous spectrum of the classical master equation for $\prm=3$ and $N=128$. Notice the two quasi-degenerate eigenvalues
(exponentially close for $T>0$ and separated by a power-law gap for $T=0$), which effectively freeze the imaginary-time
filtering capability of the master equation. 
In the thermodynamic limit the system remains ``stuck'' in the paramagnetic minimum around $m=0$.}
\label{fig:res_P3_SA}
\end{figure*}
%
Let us now turn to the crucial question: is SA better or worse than QA-RT for $\prm\ge 3$, where a first-order transition is present?
The residual energy data when $T_{\rmf}=0$, shown in Fig.~\ref{fig:res_P3_SA}(a), display now a considerable size-dependence, 
at variance with the $\prm=2$ case, and are well described by an exponential relaxation of the form
\begin{equation}
\epsilon^{\res}_N(\tau) \simeq \frac{1}{2} \nep^{-\tau/\tau^*_N} \;,
\end{equation}
where the characteristic time-scale $\tau^*_N$ increases now as a {\em power-law} of $N$, for instance
$\tau^*_N\sim \sqrt{N}$ for $\prm=3$, see Fig.~\ref{fig:res_P3_SA}(b).
Notice that the pre-factor in front of the exponential is very close to $1/2$, with a negligible size dependence, 
at variance with the $C_{\prm}/N$ pre-factor appearing in the LZ-regime of QA-RT. 
To explain these features, one apply Kramer's theory to describe the escape of the probability
from the paramagnetic free-energy minimum at $T\gg T_c$.
The equilibrium free-energy density $f(m,T)$ of the classical $p$-spin model is simply given by
$f(m,T)=-(J/2) m^{\prm} - T s(m)$ where the entropy density $s(m)$ originates from the binomial coefficients and is given by
\begin{equation} 
s(m) = \log 2-{\textstyle \frac{(1-m)}{2} \log(1-m)} - {\textstyle \frac{(1+m)}{2}\log(1+m)} \;. \nonumber
\end{equation}
Fig.~\ref{fig:res_P3_SA}(c) shows $f(m,T)$ for $\prm=3$ at different temperatures above and below $T_c$.
The presence of a free energy barrier, separating the ferromagnetic and paramegnetic phases when $0<T<T_c$,
is visible also in the spectrum of the transition matrix.
Indeed the two lowest eigenvalues are very close (exponentially) even for temperature well below the transition point, 
while the first excited state is almost identical to the high temperature equilibrium state, as shown in Fig.~\ref{fig:res_P3_SA}(d).
This suggests that the dynamics is effectively described by a two-level approximation in which one considers the distribution as peaked in either one 
of the two minima of $f(m,T)$. 
Let us call $P_0(t)$ the probability that $\Pm(m,t)$ is inside the paramagnetic valley around $m=0$ 
at time $t$ (formally, we should sum over all the $m$ values inside the paramagnetic valley, say below a given value
$m_B$ marking the barrier point), and $P_1(t)$ the probability that the system is close to the ferromagnetic minimum
at or near $m=1$. Before the transition temperature $T_c$ is reached, $P_1(t)$ is zero (or even undefined, if the minimum
is not formed), while for $T<T_c$ $P_1(t)$ starts growing, but the transition rate to return back to the (wrong) paramagnetic
minimum becomes increasingly small. 
All in all, we are justified in writing a master equation for $P_0(t)$ alone in the Kramer's escape form:
\begin{equation}\label{eq:rates}
\frac{d}{d t} P_0(t)\simeq -A \; \nep^{-\frac{N \Delta f(t)}{k_B T(t)} } \; P_0(t) \;,
\end{equation}
where $N\Delta f(t)$ is the free energy barrier separating the paramagnetic and the ferromagnetic minima for $T(t)<T_c$,
while the rate $A$ can be taken to be a constant (to within leading exponentials). 
Given $f(m,T)$ we can calculate the free-energy barrier $\Delta f(t)$. 
When $T$ is small, more precisely for $T\to 0$, the position of the barrier maximum $m_B$ is close to $m=0$, 
and, by expanding the entropy $s(m)$, one can easily show that 
\[ m_B \sim \left( \frac{2T}{Jp} \right)^{\frac{1}{(p-2)}} \;. \] 
Hence the barrier height $\Delta f(t) = f(m_B,T(t)) - f(0,T(t))$ can be approximated, in the same regime, as:
\begin{equation}
\Delta f (t) = \frac{J(\prm-1)}{2} 
\left( \frac{2k_B T(t)}{J \prm} \right)^{\frac{\prm}{\prm-2}} \;. 
\end{equation}
%
%
We need to solve Eq.~\eqref{eq:rates} with the initial conditions $T(0)=T_c$, $P_0(0)=1$ 
and the usual linear schedule for the temperature annealing, see Eq.~\eqref{eqn:annQC}. 
(The initial evolution for $T(t)>T_c$ can be neglected. since the system is already in the stable minimum and it 
remains always close to the equilibrium distribution.)
To do that, it is useful to use directly the temperature as an independent variable, with the substitution 
$\frac{d}{d t}\ \rightarrow \ -\frac{T_c}{\tau}\frac{d}{d T}$, which allows us to rewrite Eq.~\eqref{eq:rates} as
\begin{equation}
\frac{d}{d T}\log (P_0)=\frac{A \tau}{T_c} \nep^{-N\left(\frac{p-1}{p}\right) \left(\frac{2T}{J\prm}\right)^{\frac{2}{\prm-2}}}\ .
\end{equation}
Integrating between $T_c$ and $T_{\rmf}=0$ we get:
\begin{equation}
\log(P_0(T=0))= -\frac{A \tau}{T_c\ N^{\frac{\prm -2 }{2}}} \int_0^{T_cN^{\frac{\prm -2 }{2}}} \hspace{-5mm} \ud y \;
\nep^{-\left(\frac{p-1}{p}\right) \left(\frac{2y}{J\prm}\right)^{\frac{2}{\prm-2}}} \;, \nonumber
\end{equation}
where we made the substitution $N\ T^{\frac{2}{\prm-2}}=y^{\frac{2}{\prm-2}}$ in order to eliminate the dependence on 
$N$ from the integrand.
Notice that for any $\prm\ge 3$ the integrand is a function that drops exponentially fast to zero, 
so that we can send the upper integration limit to infinity, making a negligible error of the order $O(\nep^{-N})$:
\begin{equation} \label{eqn:P0_int}
\log(P_0(\tau))\simeq -\frac{A \tau}{T_c\ N^{\frac{\prm -2 }{2}}} \int_0^{\infty} \hspace{-2mm} \ud y \;
\nep^{-\left(\frac{p-1}{p}\right) \left(\frac{2y}{J\prm}\right)^{\frac{2}{\prm-2}}} \;.
\end{equation}
Therefore we predict that the final excitation probability behaves as 
\begin{equation}
P_0(\tau)\simeq \nep^{-\tau/\tau^*_N} \;,
\end{equation}
with a characteristic time $\tau^*_N$ increasing {\em polynomially} with the system size $N$, as
\begin{equation}
\tau^*_N =\frac{T_c}{\tilde{A}} \; N^{\frac{\prm -2}{2}} \;,
\end{equation}
where $\tilde{A}$ contains both the factor $A$ and the contribution from the integral in Eq.~\eqref{eqn:P0_int}.
This prediction is in very good agreement with our numerical data, see Fig.~\ref{fig:res_P3_SA}(b). 
%
The final outcome of this calculation suggests that in the thermodynamic limit $N\to\infty$ the system is trapped
in the paramagnetic minimum, i.e. $P_0(\tau)\sim 1$, hence the residual energy per-spin remains stuck at the value $\simeq 1/2$:
\begin{equation} \label{eqn:eres_SA_thermolimit}
\epsilon^{\res}_{N\to \infty}(\tau) \simeq \frac{1}{2} \;. 
\end{equation} 

\subsection{$p=\infty$: The Grover limit}  \label{sec:grover}
%
Some final remarks are necessary to explain the connection with the Grover problem, i.e. $\prm({\rm odd})=\infty$. 
If we apply our analysis to this situation, we find that both QA-RT --- without an optimized schedule ---  and SA are characterized 
by a time scale that diverges as $2^N$, in analogy with what is known in the literature~\cite{Grover_PRL97,Bapst_JSTAT12}. 
Therefore for $N\to \infty$ none of the two processes is able to attain the correct minimal energy configuration. 
While this is clear from our SA analysis --- indeed, our arguments suggest that SA is stuck even for finite $\prm\ge 3$ ---, 
the conclusion it is slightly more subtle for QA-RT. 
Here it is important to notice that the logarithmically decreasing envelope in Eq.~\eqref{eqn:eres_qa_rt_p3} originates 
essentially from the fact that the LZ form $\epsilon^{\LZ}_N(\tau) = \frac{C_{\rm p}}{N} \nep^{-\tau/\tau^*_N}$ has
a characteristic $1/N$ pre-factor, since the constant $C_{\rm p}\simeq \Delta(\Gamma=0) \approx J \prm$ is essentially
the (finite) excitation energy that the system reaches when the important LZ transition is precisely that occurring at
the critical point. The fact that $C_{\rm p}$ is {\em not extensive} with $N$ is crucial. 
This condition, however, is valid only for {\em finite} $\prm$. 
For $\prm ({\rm odd}) = \infty$, indeed, the system has only three eigenvalues when $\Gamma=0$: 
$0$, which is massively degenerate, $2^{N-1}$-fold, and $\pm N/2$.
Hence $\Delta(\Gamma=0) \propto N$ and the envelope will no longer be a logarithmically decreasing function of $\tau$. 
In conclusion, even QA is ``stuck'' and guaranteed to fail for $N\to \infty$ if we {\em first} send $\prm\to \infty$:
the possible battle against a classical algorithm is on the time-scale needed to reach the solution for {\em finite} $N$.
And here a QA schedule optimization \cite{Roland_PRA02} can lead to the celebrated Grover quadratic speed-up \cite{Grover_PRL97}.  

\section{Results: QA in real-time versus imaginary-time} \label{sec:results_IT}

Let us now consider the case of the imaginary-time QA Schr\"odinger dynamics, to contrast it to the physical real-time QA. 
The behaviour of the two dynamics turns out to be very different, as already found in the Ising chain case \cite{Zanca_2016}.  
In particular, the closing of the gap at the transition point is highly irrelevant to the QA-IT dynamics: as the gap {\em re-opens} and stays finite
in the whole ferromagnetic phase with $\Gamma<\Gamma_c$, the dynamics will filter very effectively the instantaneous ground
state; the scaling of the residual energy can be easily explained through the adiabatic theorem for imaginary-time processes\cite{Morita_JMP08,DeGrandi_PRB11}, 
which predicts an asymptotic $1/\tau^2$ behaviour. 
%
\begin{figure}[ht!]
\begin{center}
\includegraphics[width=80mm]{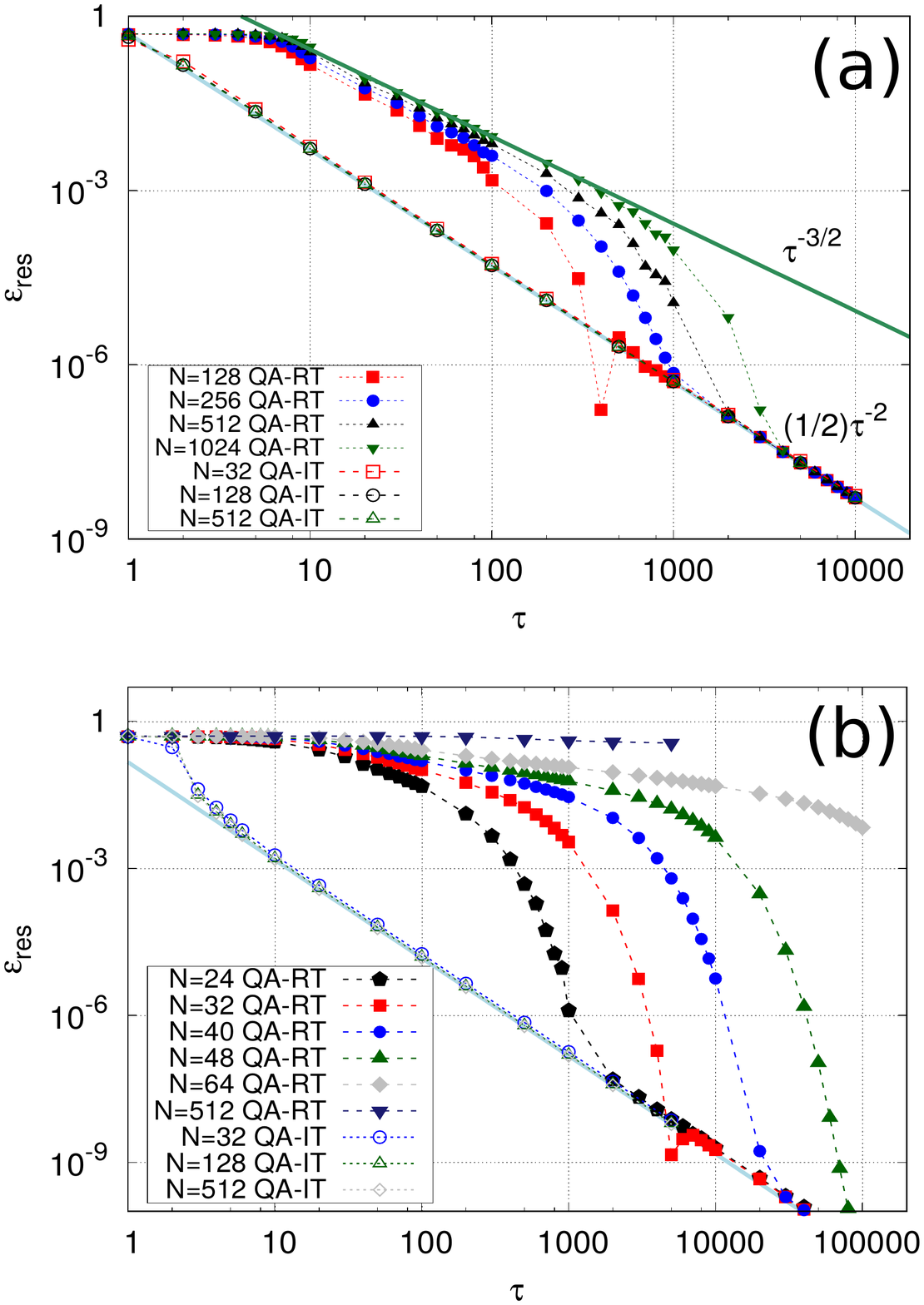}
\end{center}
\caption{(a) The residual energy density $\epsilon^{\res}_N(\tau)$ {\it vs} annealing time for QA-RT for $\prm=2$,
compared to the results obtained through an imaginary-time Schr\"odinger dynamics (QA-IT).    
Here $\Gamma_{\rmi}=2$.
(b) Same as in (a), but for $\prm=3$. 
}
\label{fig:RT_IT}
\end{figure}
%
Indeed, as predicted in Ref.~\onlinecite{Morita_JMP08}, and clearly visible from the numerical results in Fig. \ref{fig:RT_IT}, the two QA dynamics share the 
same asymptotic regime: the only caveat is that while QA-IT obeys it without restrictions, 
QA-RT does it only for $\tau\gg \Delta_N^{-2}$, hence only for finite $N$.  
For QA-IT, the predictions of the adiabatic theorem \cite{Morita_JMP08,DeGrandi_PRB11} imply that for essentially all values of $\tau$
\begin{equation} \label{eqn:eres_ITadiabatic}
\epsilon^{\res}(\tau) \simeq \frac{\Gamma_{\rmi}^2}{p^3} \frac{1}{\tau^2} \;, 
\end{equation}
with negligible size-corrections.   

\section{Discussion and conclusions} \label{sec:conclusions}
%
In this work we compared classical thermal annealing (SA) and quantum annealing  (QA) --- both in real time and imaginary time --- on a fully-connected $p$-spin Ising ferromagnet.
Thanks to the mean-field character of the model, and a permutation symmetry, we were able to solve exactly the dynamical equations for quite large systems, $N\sim 10^3$ spins.
It was then possible to perform a careful finite-size-scaling analysis to extract the relevant behaviour for large annealing times and in the thermodynamic limit $N\rightarrow \infty$.
Our results show a remarkable difference in the performance of the different annealing strategies, depending on the order of the transition: second ($\prm =2$) or first ($\prm\ge 3$) order.
In particular we found an {\em exponential} speedup in SA with respect to QA when the system crosses a second-order phase transition. 
For first-order phase transition instead, SA becomes less and less efficient for increasing system size, until it remains stuck in the disordered phase when the thermodynamic limit is approached.
To find the large annealing time behaviour when $N\rightarrow \infty$ in QA-RT dynamics,
we developed a novel approach based on the analysis of the geometric envelope of finite-size data in a range of annealing time where 
the system evolution can be effectively described by the interference of the two lowest energy levels (Landau-Zener regime).
For $\prm\ge 3$, this analysis predicts a slow decay of the residual energy with the annealing time $\tau$, with an inverse-logarithmic behaviour $1/\log(\tau)$, which means a limited quantum speedup as defined in Ref.\onlinecite{Ronnow2014}.  We would emphasize that the present conclusion has been drawn by direct solutions of the Schr\"odinger and master equations in combination with a non-equilibrium analysis based on the Landau-Zener formula, whereas most studies so far have been based on numerical simulations, a notable exception being the one-dimensional case discussed in  Ref.~\onlinecite{Zanca_2016}.  The envelope method is hence particularly useful to study the asymptotic regime of systems that cross a first-order phase transition, where usually one can solve the dynamics 
only for small values $N$ and $\tau$. 


A possible line of investigation, which we leave to future work, concerns the study of the open-system quantum dynamics of the same model, to elucidate the competition 
between thermal effects due to the environment, and genuine quantum tunnelling effects as discussed recently in Ref.\onlinecite{Jiang2017}.

\section*{Acknowledgements}
We acknowledge fruitful discussions with A. Parola and A. Scardicchio. RF kindly acknowledges support from the National Research Foundation of Singapore
(CRP - QSYNC), the Oxford Martin School and EU through the project QUIC. GES acknowledges support from EU through ERC MODPHYSFRICT.
Part of the work of HN has been supported by the KAKENHI Grant No. 26287086 by the Japan Society for the Promotion of Science.
\appendix
\section{The envelope construction to finite-$N$ annealing data} \label{app:envelope}
\begin{figure}[ht!]
\begin{center}
\includegraphics[width=80mm]{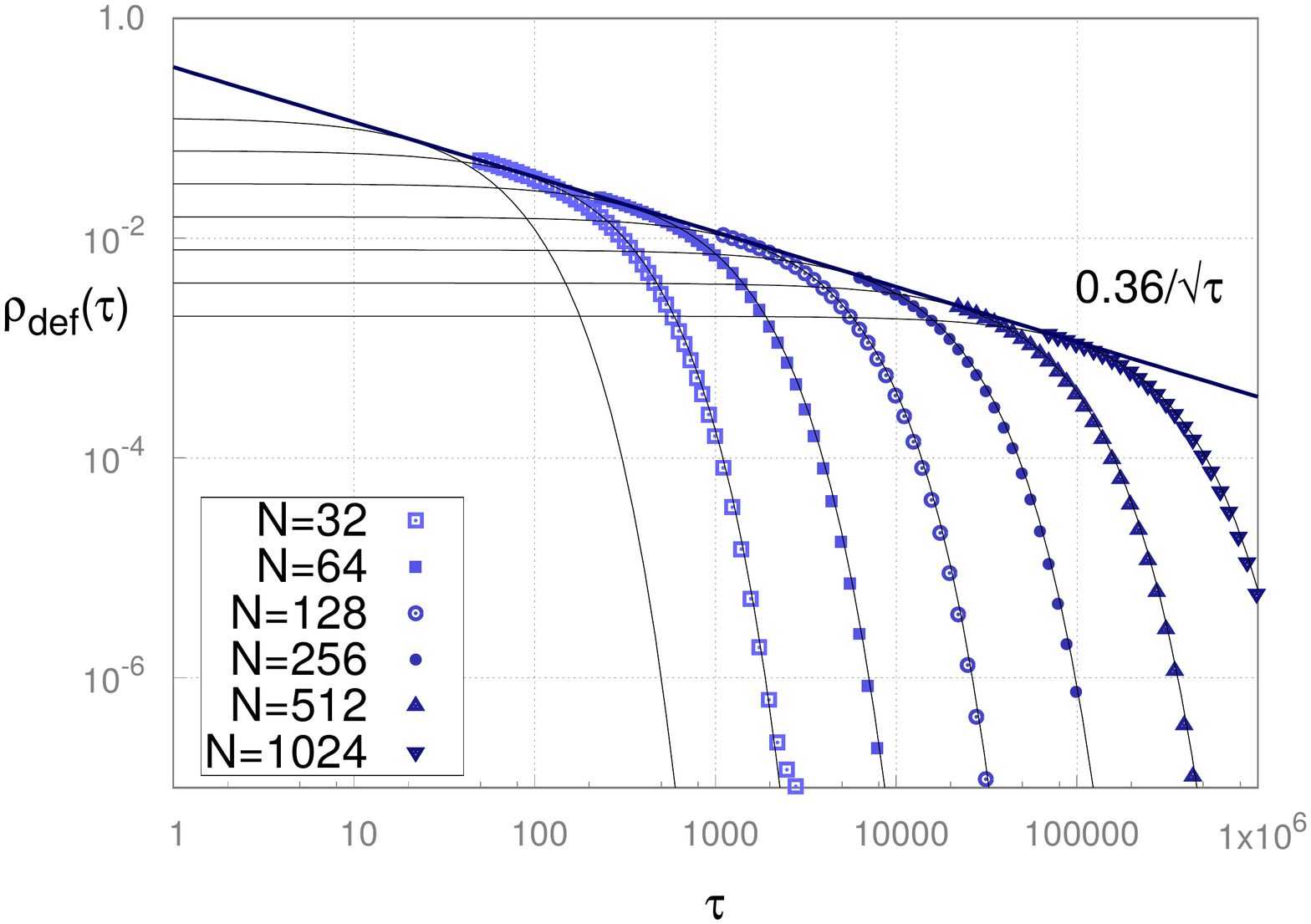}
\caption{
The envelope construction illustrated for the transverse-field Ising chain.
}
\label{fig:envelope}
\end{center}
\end{figure}

The strategy is based on the geometrical construction of the {\em envelope} of all the finite-$N$ LZ-curves.
Briefly, if $f_{u}(x)$ is a family of functions which smoothly depends on some parameter $u$, 
its envelope $e(x)$ is a function which is tangent at each point to a member of the family, a condition that is enforced
by solving $\partial_u f_u(x) = 0$ to find $u(x)$ and then setting 
\begin{equation}
e(x) = f_{u(x)}(x) \;.
\end{equation}
If we now identify $u\mapsto N$ (which we assume to be real, rather than integer), $x\mapsto \tau$, and 
$f_u(x)\mapsto \epsilon^{\LZ}_N(\tau)$, we can construct the geometrical envelope of these finite-$N$ Landau-Zener annealing data.
We illustrate in Fig., \ref{fig:envelope} how this procedure works in the transverse-field Ising chain, where the well-known Kibble-Zurek form of
the asymptotic residual energy is $\epsilon^{\res}(\tau)\sim 1/\sqrt{\tau}$. 

\section{Mapping of a classical master equation into an imaginary time quantum problem} \label{app:mapping}
%
%
%

Let us comment here briefly on the reason why we did not adopt the symmetrization strategy of Ref.~\onlinecite{Nishimori_PRE14}
to transform our SA master equation into an effective imaginary-time Schr\"odinger problem.
The reason is that, as pointed out in Ref.~\onlinecite{Nishimori_PRE14}, to properly perform the symmetrization
when the temperature depends on time one would need to account, in the effective quantum Hamiltonian, 
for an extra potential term of the form:
%
\begin{equation} \label{V_sigma_anneal:eqn}
V^{\rm neq}_{\sigma} =- \frac{1}{2} {\dot \beta} \Big( H_{\rm C}(\sigma) - \langle H_{\rm C} \rangle_{\rm eq} \Big) \;,
\end{equation}
which originates from the time derivative of the equilibrium Boltzmann distribution.
As it turns out, we have verified that this term {\em cannot} be neglected in the present annealing set-up: 
the price for that would be an unphysical violation of the total probability conservation. 
In the end, the mapping to an effective quantum problem, if properly pursued with due account of this extra term, does not add 
any real advantage to the more conventional strategy of working directly with a master equation for $\Pm(m,t)$
(indeed, we found that the ensuing equations are numerically less stable). 

Nevertheless, the analogy of the classical master equation with an imaginary-time Schr\"odinger problem is quite
inspiring,
and can be made precise, as already mentioned, by a suitable symmetrization of the transition rate matrix $W_{\sigma,\sigma'}$.
This correctly suggests, for instance, that the dynamics proceeds by ``filtering-out'' the higher excited eigenstates from the time
evolving $P(\sigma,t)$.
%

\section{Adiabatic approximation for a classical master equation} \label{app:adiab}
Here we present a modified version of the adiabatic expansion, which is suited to describe the solution of a classical master equation;
essentially, we are interested in the evolution in time of a probability distribution instead of a quantum wave function.

We write the master equation in the following form:
\begin{equation}\label{eq:ME_k}
-\frac{\partial }{\partial t} P(\sigma,t) = \sum_{\sigma'} K_{\sigma,\sigma'} P(\sigma',t);\,
\end{equation}
where $\sigma=(\sigma^1,\dots,\sigma^N)$ is a configuration of $N$ Ising variables and $K_{\sigma,\sigma'}$ a stochastic
matrix \cite{vanKampen:book}, $\sum_{\sigma} K_{\sigma,\sigma'}=0$, which ensures probability conservation. 
Since we want to use Eq.~(\ref{eq:ME_k}) to describe a thermal process, we impose that the thermal distribution 
$P^{\rm eq}(\sigma)=\nep^{-\beta H_{\rm C}(\sigma)}/Z$ is the equilibrium state, i.e.,
$\sum_{\sigma'} K_{\sigma,\sigma'} P^{\rm eq}(\sigma')=0$.
$P^{\rm eq}(\sigma')$ is therefore a {\em right} eigenvector of $K_{\sigma,\sigma'}$ with null eigenvalue $E_0=0$. 
The {\em left} eigenstate $\tilde{P}_0(\sigma)$ corresponding to $E_0=0$ is the row vector with all elements equal to 1.

In general $K_{\sigma,\sigma'}$ has a basis of right eigenvectors $P_n(\sigma)$ and left eigenvectors $\tilde{P}_n(\sigma)$ 
which satisfy the following equations
\begin{eqnarray}
\sum_{\sigma'} K_{\sigma,\sigma'} P_n(\sigma') &=& E_n P_n(\sigma)\;,\\
\sum_{\sigma} \tilde{P}_n(\sigma) K_{\sigma,\sigma'} &=& E_n \tilde{P}_n(\sigma')\;,\\
\sum_{\sigma} \tilde{P}_m(\sigma) P_n(\sigma) &=&\delta_{m,n} \;.
\end{eqnarray}
To study the spectrum of the operator $K_{\sigma,\sigma'}$ it is useful to apply a standard symmetrization procedure\cite{vanKampen:book} to map the
transition rate matrix $K_{\sigma,\sigma'}$ into a symmetric operator $\calH_{\sigma,\sigma'}$.
The new eigenstates are related to the right and left eigenvectors of $K_{\sigma,\sigma'}$ by
\begin{eqnarray}
\sum_{\sigma'} \calH_{\sigma,\sigma'} \phi_n(\sigma') &=& E_n \, \phi_n(\sigma)\;, \nonumber \\
P_n(\sigma) &=& \sqrt{P^{\rm eq}(\sigma)} \, \phi_n(\sigma) \nonumber \\
\tilde{P}_n(\sigma) &=& \frac{1}{\sqrt{P^{\rm eq}(\sigma)}} \, \phi_n(\sigma) \;.
\end{eqnarray}
Now let us focus on Eq.~(\ref{eq:ME_k}), when it describes a process with a time dependent temperature $T(t)$.
It is possible to expand the probability $P(\sigma,t)$ on the right eigenvector basis of $K_{\sigma,\sigma'}$
\begin{equation}
P(\sigma,t)=\sum_{n} c_n(t) \, P^{T(t)}_n(\sigma) \; \nep^{\int_t^{\tau} E_n(t') \ud t'} \;,
\end{equation}
where the dependence of the vectors $P_n$ and the eigenvalues $E_n$ on time comes from $T(t)$. 
Henceforth we will not write it explicitly. 
Eq.~\eqref{eq:ME_k} thus becomes 
\begin{eqnarray}
-\frac{\partial}{\partial t}P(\sigma,t) &=&
-\sum_{n} \nep^{\int_t^{\tau}E_n} \left[ \dot{c}_n P_n(\sigma) + 
c_n \partial_t  P_n(\sigma) \right. \nonumber \\
&& \hspace{20mm} \left. - E_n P_n(\sigma) \right] \nonumber \\
&=& \sum_n c_n(t) \, E_n \, P_n(\sigma)\, \nep^{\int_t^{\tau}E_n \ud t'}\;.
\end{eqnarray}
The exponential term is written so that the "trivial" diagonal evolution cancels, 
leaving an equation for the coefficients $c_n(t)$. After some steps identical to those exploited in the usual quantum adiabatic theorem \cite{Morita_JMP08}, we obtain
\begin{equation}\label{eq:eq_k}
\dot{c}_m(t)+\sum_n c_n(t) \left(\tilde{P}_m| \partial_t P_n \right)\nep^{\int_t^{\tau}(E_n-E_m) \ud t'}=0 \;,
\end{equation}
where $ (\ |\ )$ is the scalar product in the space of configurations $\sigma$.
Now assume that the system at $t=0$ is in the thermal equilibrium state, 
so that we can neglect all other contributions in the sum in Eq. (\ref{eq:eq_k}). This is accomplished by setting $c_n= \delta_{n,0}$. 
Therefore we write 
\begin{equation}\label{eq:eq_k_appr}
-\dot{c}_m(t)\simeq\left(\tilde{P}_m(t)| \partial_t P^{\rm eq}(t) \right)\nep^{-\int_t^{\tau}\Delta_{m0} \ud t'} \;,
\end{equation}
exploiting the fact that $P_0(\sigma,t)=P^{\rm eq}(\sigma,t)$ and defining $\Delta_{m0}=E_m-E_0=E_m$.
The next step is to change variable in the integral over the gap, in order to extract the adiabatic parameter $\dot{T}$. If the temperature is a linear decreasing function of time,  its derivative can be written as $-T_{\rmi}/\tau$, where $\tau$ is the total time of the process and $1/\tau$ will be used as the ``small'' parameter for the power series expansion
\begin{equation}
\nep^{-\int_t^{\tau}\Delta_{m0} \ud t'}=\nep^{-\frac{\tau}{T_{\rmi}}\int_{T_{\rmf}}^T \Delta_{m0} \ud T'}\;.
\end{equation}
Eq.~\eqref{eq:eq_k_appr} can be solved by an integration by part to obtain the leading order in powers of $1/\tau$
\begin{equation}\label{eq:adia_c_sol1}
c_m(\tau) \simeq - \frac{T_{\rmi}}{\tau} \left[ B(T_{\rmf})-B(T_{\rmi})\nep^{-\frac{\tau}{T_{\rmi}}\int_{T_{\rmf}}^{T_{\rmi}} \Delta_{m0} \ud T'}\right] +O\left(\frac{1}{\tau^2}\right),
\end{equation}
where the temperature coefficients $B(T)$ are defined through
$B(T)=\frac{\left(\tilde{P}_m| \partial_{\beta} P^{\rm eq} \right)}{\Delta_{m0}}$,
evaluated at temperature $T$.
Since the exponential term in Eq.~\eqref{eq:adia_c_sol1} adds a vanishing contribution for large values of $\tau$, 
the solution can be further approximated as
\begin{equation}\label{eq:adia_c_sol1.2}
c_m(\tau) \simeq - \frac{T_{\rmi}}{\tau} \frac{\left(\tilde{P}_m(T) | \partial_{\beta} P^{\rm eq}(T) \right)_{T=T_{\rmf}} }{\Delta_{m0}} \;.
\end{equation}

The difficulty in applying the adiabatic expansion to a master equation lies in the evaluation of the term 
$\left(\tilde{P}_m(T)| \partial_{\beta} P^{\rm eq}(T) \right)$.
This is more easily accomplished by means of the symmetrized eigenstates. 
We start by noticing that
\begin{equation}
\partial_{\beta} P^{\rm eq}(\sigma)=-P^{\rm eq}(\sigma)\left(H_{\rm C}(\sigma) - \langle H_{\rm C} \rangle^{\rm eq} \right)\;.
\end{equation}
The term with the average energy is hence neglected because it is proportional to the identity matrix and therefore it gives no contribution inside the scalar product.
Using the relation between the left eigenvector of $K_{\sigma,\sigma'}$ and the eigenstate of $\calH$, we can write
\begin{eqnarray}\label{eq:scalar1}
\left(\tilde{P}_m| \partial_{\beta} P^{\rm eq}\right) &=&
-\sum_{\sigma} \frac{\phi_m(\sigma)}{\sqrt{P^{\rm eq}(\sigma)}}P^{\rm eq}(\sigma) H_{\rm C}(\sigma) \nonumber \\
&=& -\langle \phi_m|H_{\rm C}|\phi_0\rangle\;. 
\end{eqnarray}
On the other hand, if we substitute directly the eigenvectors $\phi_m$ instead of $P_m$ and $\tilde{P}_m$, we obtain
\begin{eqnarray}
\partial_{\beta} P^{\rm eq}(\sigma) &=& \partial_\beta (\sqrt{P^{\rm eq}(\sigma)} \phi_0(\sigma)) \nonumber \\
&=& -\frac{1}{2}\left(H_{\rm C} - \langle H_{\rm C} \rangle^{\rm eq} \right)\phi_0(\sigma) + 
\sqrt{P^{\rm eq}} \partial_{\beta}\phi_0(\sigma)\;. \nonumber
\end{eqnarray}
So the scalar product $\left(\tilde{P}_m| \partial_{\beta} P^{\rm eq}\right)$ can be written as
\begin{equation}\label{eq:scalar2}
\left(\tilde{P}_m| \partial_{\beta} P^{\rm eq}\right)= -\frac{1}{2}\langle \phi_m|H_{\rm C}|\phi_0\rangle + \langle \phi_m|\partial_{\beta}\phi_0\rangle \;. 
\end{equation}
Comparing  Eqs.~\eqref{eq:scalar1} and \eqref{eq:scalar2}, one immediately notices that the term 
$-\langle \phi_m|H_{\rm C}|\phi_0\rangle $ appears in both, which implies that: 
\begin{equation}
-\frac{1}{2}\langle \phi_m|H_{\rm C}|\phi_0\rangle =\langle \phi_m|\partial_{\beta}\phi_0\rangle\;,
\end{equation}
so that our final expression becomes
\begin{equation}
\left( \tilde{P}_m| \partial_{\beta} P^{\rm eq} \right)=
 2 \langle \phi_m| \partial_{\beta} \phi_0 \rangle=
-2 \frac{\langle \phi_m| \partial_{\beta} \calH |\phi_0\rangle}{\Delta_{m0}}  \;.
\end{equation}
Therefore it finally is possible to write the 
asymptoticg solution for the coefficients $c_m(\tau)$ using the effective Hamiltonian $\calH$, 
at a generic final temperature $T_{\rm f}$
\begin{equation}\label{eq:adia_c_sol2}
c_m(\tau) \simeq \frac{2 T_{\rmi}}{\tau} \; \frac{\langle \phi_m| \partial_{\beta} \calH |\phi_0 \rangle_{T=T_{\rmf}}}{\Delta^2_{m0}} \;.
\end{equation}
The adiabatic expansion for the solution of a classical master equation leads to a final expression analogous to what is obtained for quantum IT dynamics\cite{Morita_JMP08}.

Although we found numerically more convenient to solve directly the Master Equation, 
the symmetrized form of the transition matrix $W_{\sigma,\sigma'}$, 
{\it i.e.}, the effective ``quantum'' Hamiltonian $\calH$, is useful to study the behaviour of the adiabatic solution.
After the permutation symmetry is exploited to describe the system in terms of its magnetization $m$, the effective Hamiltonian reads
\begin{equation}\label{eq:Heff_hb}
\calH_{m,m'}=
\begin{cases}
\frac{N}{2}\sum_{\alpha=\pm}  \frac{(1-\alpha m)}{1+\nep^{\frac{\beta}{2}\Delta E_\alpha}} &
m'=m \;, \vspace{2mm} \\
-\frac{N}{2}\frac{ \sqrt{1-m^2+\frac{2}{N}}}{2\cosh \left( \frac{\beta}{2}\Delta E_+\right)}&
 m'=m+1 \;.
\end{cases}
\end{equation}
In the adiabatic expansion $\calH_{m,m'}$ appears through its inverse-temperature derivative, which is
\begin{equation}\label{eq:Heff_hb_der}
\partial_{\beta} \calH_{m,m'}=
\begin{cases}
-\frac{N}{4}\sum_{\alpha=\pm}  \frac{(1-\alpha m)\Delta E_{\alpha} \nep^{\frac{\beta}{2}\Delta E_\alpha}}{\left(1+\nep^{\frac{\beta}{2}\Delta E_\alpha}\right)^2} &
 m'=m \; ,\\
\frac{N}{4}\frac{ \sqrt{1-m^2+\frac{2}{N}}  \Delta E_+ \tanh \left( \frac{\beta}{2}\Delta E_+\right)}{2\cosh \left( \frac{\beta}{2}\Delta E_+\right)}&
 m'=m+1 \;.
\end{cases}
\end{equation}
It is easy to see that if one takes the limit $\beta \rightarrow \infty$ ($T\rightarrow 0$) all the matrix elements $\calH_{m,m'}$ vanish exponentially, whence 
the absence of an adiabatic expansion in powers of $\frac{1}{\tau}$ when $T_{\rmf}=0$ in an annealing process.


\begin{thebibliography}{61}
\expandafter\ifx\csname natexlab\endcsname\relax\def\natexlab#1{#1}\fi
\expandafter\ifx\csname bibnamefont\endcsname\relax
  \def\bibnamefont#1{#1}\fi
\expandafter\ifx\csname bibfnamefont\endcsname\relax
  \def\bibfnamefont#1{#1}\fi
\expandafter\ifx\csname citenamefont\endcsname\relax
  \def\citenamefont#1{#1}\fi
\expandafter\ifx\csname url\endcsname\relax
  \def\url#1{\texttt{#1}}\fi
\expandafter\ifx\csname urlprefix\endcsname\relax\def\urlprefix{URL }\fi
\providecommand{\bibinfo}[2]{#2}
\providecommand{\eprint}[2][]{\url{#2}}

\bibitem[{\citenamefont{Lucas}(2014)}]{Lucas2014}
\bibinfo{author}{\bibfnamefont{A.}~\bibnamefont{Lucas}},
  \bibinfo{journal}{Frontiers in Phys.} \textbf{\bibinfo{volume}{2}},
  \bibinfo{pages}{1} (\bibinfo{year}{2014}).

\bibitem[{\citenamefont{Finnila et~al.}(1994)\citenamefont{Finnila, Gomez,
  Sebenik, Stenson, and Doll}}]{Finnila_CPL94}
\bibinfo{author}{\bibfnamefont{A.~B.} \bibnamefont{Finnila}},
  \bibinfo{author}{\bibfnamefont{M.~A.} \bibnamefont{Gomez}},
  \bibinfo{author}{\bibfnamefont{C.}~\bibnamefont{Sebenik}},
  \bibinfo{author}{\bibfnamefont{C.}~\bibnamefont{Stenson}}, \bibnamefont{and}
  \bibinfo{author}{\bibfnamefont{J.~D.} \bibnamefont{Doll}},
  \bibinfo{journal}{Chem. Phys. Lett.} \textbf{\bibinfo{volume}{219}},
  \bibinfo{pages}{343} (\bibinfo{year}{1994}).

\bibitem[{\citenamefont{Kadowaki and Nishimori}(1998)}]{Kadowaki_PRE98}
\bibinfo{author}{\bibfnamefont{T.}~\bibnamefont{Kadowaki}} \bibnamefont{and}
  \bibinfo{author}{\bibfnamefont{H.}~\bibnamefont{Nishimori}},
  \bibinfo{journal}{Phys. Rev. E} \textbf{\bibinfo{volume}{58}},
  \bibinfo{pages}{5355} (\bibinfo{year}{1998}).

\bibitem[{\citenamefont{Brooke et~al.}(1999)\citenamefont{Brooke, Bitko,
  Rosenbaum, and Aeppli}}]{Brooke_SCI99}
\bibinfo{author}{\bibfnamefont{J.}~\bibnamefont{Brooke}},
  \bibinfo{author}{\bibfnamefont{D.}~\bibnamefont{Bitko}},
  \bibinfo{author}{\bibfnamefont{T.~F.} \bibnamefont{Rosenbaum}},
  \bibnamefont{and} \bibinfo{author}{\bibfnamefont{G.}~\bibnamefont{Aeppli}},
  \bibinfo{journal}{Science} \textbf{\bibinfo{volume}{284}},
  \bibinfo{pages}{779} (\bibinfo{year}{1999}).

\bibitem[{\citenamefont{Santoro et~al.}(2002)\citenamefont{Santoro,
  {Marto\v{n}\'{a}k}, Tosatti, and Car}}]{Santoro_SCI02}
\bibinfo{author}{\bibfnamefont{G.~E.} \bibnamefont{Santoro}},
  \bibinfo{author}{\bibfnamefont{R.}~\bibnamefont{{Marto\v{n}\'{a}k}}},
  \bibinfo{author}{\bibfnamefont{E.}~\bibnamefont{Tosatti}}, \bibnamefont{and}
  \bibinfo{author}{\bibfnamefont{R.}~\bibnamefont{Car}},
  \bibinfo{journal}{Science} \textbf{\bibinfo{volume}{295}},
  \bibinfo{pages}{2427} (\bibinfo{year}{2002}).

\bibitem[{\citenamefont{Santoro and Tosatti}(2006)}]{Santoro_JPA06}
\bibinfo{author}{\bibfnamefont{G.~E.} \bibnamefont{Santoro}} \bibnamefont{and}
  \bibinfo{author}{\bibfnamefont{E.}~\bibnamefont{Tosatti}},
  \bibinfo{journal}{J. Phys. A: Math. Gen.} \textbf{\bibinfo{volume}{39}},
  \bibinfo{pages}{R393} (\bibinfo{year}{2006}).

\bibitem[{\citenamefont{Morita and Nishimori}(2008)}]{Morita_JMP08}
\bibinfo{author}{\bibfnamefont{S.}~\bibnamefont{Morita}} \bibnamefont{and}
  \bibinfo{author}{\bibfnamefont{H.}~\bibnamefont{Nishimori}},
  \bibinfo{journal}{J. Math. Phys.} \textbf{\bibinfo{volume}{49}},
  \bibinfo{pages}{125210} (\bibinfo{year}{2008}).

\bibitem[{\citenamefont{Farhi et~al.}(2001)\citenamefont{Farhi, Goldstone,
  Gutmann, Lapan, Lundgren, and Preda}}]{Farhi_SCI01}
\bibinfo{author}{\bibfnamefont{E.}~\bibnamefont{Farhi}},
  \bibinfo{author}{\bibfnamefont{J.}~\bibnamefont{Goldstone}},
  \bibinfo{author}{\bibfnamefont{S.}~\bibnamefont{Gutmann}},
  \bibinfo{author}{\bibfnamefont{J.}~\bibnamefont{Lapan}},
  \bibinfo{author}{\bibfnamefont{A.}~\bibnamefont{Lundgren}}, \bibnamefont{and}
  \bibinfo{author}{\bibfnamefont{D.}~\bibnamefont{Preda}},
  \bibinfo{journal}{Science} \textbf{\bibinfo{volume}{292}},
  \bibinfo{pages}{472} (\bibinfo{year}{2001}).

\bibitem[{\citenamefont{Kirkpatrick et~al.}(1983)\citenamefont{Kirkpatrick,
  C.~D.~Gelatt, and Vecchi}}]{Kirkpatrick_SCI83}
\bibinfo{author}{\bibfnamefont{S.}~\bibnamefont{Kirkpatrick}},
  \bibinfo{author}{\bibfnamefont{J.}~\bibnamefont{C.~D.~Gelatt}},
  \bibnamefont{and} \bibinfo{author}{\bibfnamefont{M.~P.}
  \bibnamefont{Vecchi}}, \bibinfo{journal}{Science}
  \textbf{\bibinfo{volume}{220}}, \bibinfo{pages}{671} (\bibinfo{year}{1983}).

\bibitem[{\citenamefont{Harris et~al.}(2010)}]{Harris_PRB10}
\bibinfo{author}{\bibfnamefont{R.}~\bibnamefont{Harris}} \bibnamefont{et~al.},
  \bibinfo{journal}{Phys. Rev. B} \textbf{\bibinfo{volume}{82}},
  \bibinfo{pages}{024511} (\bibinfo{year}{2010}).

\bibitem[{\citenamefont{Johnson et~al.}(2011)}]{Johnson_Nat11}
\bibinfo{author}{\bibfnamefont{M.~W.} \bibnamefont{Johnson}}
  \bibnamefont{et~al.}, \bibinfo{journal}{Nature}
  \textbf{\bibinfo{volume}{473}}, \bibinfo{pages}{194} (\bibinfo{year}{2011}).

\bibitem[{\citenamefont{R{\o}nnow et~al.}(2014)\citenamefont{R{\o}nnow, Wang,
  Job, Boixo, Isakov, Wecker, Martinis, Lidar, and Troyer}}]{Ronnow2014}
\bibinfo{author}{\bibfnamefont{T.~F.} \bibnamefont{R{\o}nnow}},
  \bibinfo{author}{\bibfnamefont{Z.}~\bibnamefont{Wang}},
  \bibinfo{author}{\bibfnamefont{J.}~\bibnamefont{Job}},
  \bibinfo{author}{\bibfnamefont{S.}~\bibnamefont{Boixo}},
  \bibinfo{author}{\bibfnamefont{S.~V.} \bibnamefont{Isakov}},
  \bibinfo{author}{\bibfnamefont{D.}~\bibnamefont{Wecker}},
  \bibinfo{author}{\bibfnamefont{J.~M.} \bibnamefont{Martinis}},
  \bibinfo{author}{\bibfnamefont{D.~A.} \bibnamefont{Lidar}}, \bibnamefont{and}
  \bibinfo{author}{\bibfnamefont{M.}~\bibnamefont{Troyer}},
  \bibinfo{journal}{Science} \textbf{\bibinfo{volume}{345}},
  \bibinfo{pages}{420} (\bibinfo{year}{2014}).

\bibitem[{\citenamefont{{Marto\v{n}\'{a}k}
  et~al.}(2002)\citenamefont{{Marto\v{n}\'{a}k}, Santoro, and
  Tosatti}}]{Martonak_PRB02}
\bibinfo{author}{\bibfnamefont{R.}~\bibnamefont{{Marto\v{n}\'{a}k}}},
  \bibinfo{author}{\bibfnamefont{G.~E.} \bibnamefont{Santoro}},
  \bibnamefont{and} \bibinfo{author}{\bibfnamefont{E.}~\bibnamefont{Tosatti}},
  \bibinfo{journal}{Phys. Rev. B} \textbf{\bibinfo{volume}{66}},
  \bibinfo{pages}{094203} (\bibinfo{year}{2002}).

\bibitem[{\citenamefont{{Marto\v{n}\'{a}k}
  et~al.}(2004)\citenamefont{{Marto\v{n}\'{a}k}, Santoro, and
  Tosatti}}]{Martonak_PRE04}
\bibinfo{author}{\bibfnamefont{R.}~\bibnamefont{{Marto\v{n}\'{a}k}}},
  \bibinfo{author}{\bibfnamefont{G.~E.} \bibnamefont{Santoro}},
  \bibnamefont{and} \bibinfo{author}{\bibfnamefont{E.}~\bibnamefont{Tosatti}},
  \bibinfo{journal}{Phys. Rev. E} \textbf{\bibinfo{volume}{70}},
  \bibinfo{pages}{057701} (\bibinfo{year}{2004}).

\bibitem[{\citenamefont{Battaglia et~al.}(2005)\citenamefont{Battaglia,
  Santoro, and Tosatti}}]{Battaglia_PRE05}
\bibinfo{author}{\bibfnamefont{D.~A.} \bibnamefont{Battaglia}},
  \bibinfo{author}{\bibfnamefont{G.~E.} \bibnamefont{Santoro}},
  \bibnamefont{and} \bibinfo{author}{\bibfnamefont{E.}~\bibnamefont{Tosatti}},
  \bibinfo{journal}{Phys. Rev. E} \textbf{\bibinfo{volume}{71}},
  \bibinfo{pages}{066707} (\bibinfo{year}{2005}).

\bibitem[{\citenamefont{Stella et~al.}(2005)\citenamefont{Stella, Santoro, and
  Tosatti}}]{Stella_PRB05}
\bibinfo{author}{\bibfnamefont{L.}~\bibnamefont{Stella}},
  \bibinfo{author}{\bibfnamefont{G.~E.} \bibnamefont{Santoro}},
  \bibnamefont{and} \bibinfo{author}{\bibfnamefont{E.}~\bibnamefont{Tosatti}},
  \bibinfo{journal}{Phys. Rev. B} \textbf{\bibinfo{volume}{72}},
  \bibinfo{pages}{014303} (\bibinfo{year}{2005}).

\bibitem[{\citenamefont{Stella et~al.}(2006)\citenamefont{Stella, Santoro, and
  Tosatti}}]{Stella_PRB06}
\bibinfo{author}{\bibfnamefont{L.}~\bibnamefont{Stella}},
  \bibinfo{author}{\bibfnamefont{G.~E.} \bibnamefont{Santoro}},
  \bibnamefont{and} \bibinfo{author}{\bibfnamefont{E.}~\bibnamefont{Tosatti}},
  \bibinfo{journal}{Phys. Rev. B} \textbf{\bibinfo{volume}{73}},
  \bibinfo{pages}{144302} (\bibinfo{year}{2006}).

\bibitem[{\citenamefont{Matsuda et~al.}(2009)\citenamefont{Matsuda, Nishimori,
  and Katzgraber}}]{Matsuda2009}
\bibinfo{author}{\bibfnamefont{Y.}~\bibnamefont{Matsuda}},
  \bibinfo{author}{\bibfnamefont{H.}~\bibnamefont{Nishimori}},
  \bibnamefont{and} \bibinfo{author}{\bibfnamefont{H.~G.}
  \bibnamefont{Katzgraber}}, \bibinfo{journal}{New J. Phys.}
  \textbf{\bibinfo{volume}{11}}, \bibinfo{pages}{073021}
  (\bibinfo{year}{2009}).

\bibitem[{\citenamefont{Young et~al.}(2010)\citenamefont{Young, Knysh, and
  Smelyanskiy}}]{Young2010}
\bibinfo{author}{\bibfnamefont{A.~P.} \bibnamefont{Young}},
  \bibinfo{author}{\bibfnamefont{S.}~\bibnamefont{Knysh}}, \bibnamefont{and}
  \bibinfo{author}{\bibfnamefont{V.~N.} \bibnamefont{Smelyanskiy}},
  \bibinfo{journal}{Phys. Rev. Lett.} \textbf{\bibinfo{volume}{104}},
  \bibinfo{pages}{020502} (\bibinfo{year}{2010}).

\bibitem[{\citenamefont{Hen and Young}(2011)}]{Hen2011}
\bibinfo{author}{\bibfnamefont{I.}~\bibnamefont{Hen}} \bibnamefont{and}
  \bibinfo{author}{\bibfnamefont{A.~P.} \bibnamefont{Young}},
  \bibinfo{journal}{Phys. Rev. E} \textbf{\bibinfo{volume}{84}},
  \bibinfo{pages}{061152} (\bibinfo{year}{2011}).

\bibitem[{\citenamefont{Farhi et~al.}(2012)\citenamefont{Farhi, Gosset, Hen,
  Sandvik, Shor, Young, and Zamponi}}]{Farhi2012}
\bibinfo{author}{\bibfnamefont{E.}~\bibnamefont{Farhi}},
  \bibinfo{author}{\bibfnamefont{D.}~\bibnamefont{Gosset}},
  \bibinfo{author}{\bibfnamefont{I.}~\bibnamefont{Hen}},
  \bibinfo{author}{\bibfnamefont{A.~W.} \bibnamefont{Sandvik}},
  \bibinfo{author}{\bibfnamefont{P.}~\bibnamefont{Shor}},
  \bibinfo{author}{\bibfnamefont{A.~P.} \bibnamefont{Young}}, \bibnamefont{and}
  \bibinfo{author}{\bibfnamefont{F.}~\bibnamefont{Zamponi}},
  \bibinfo{journal}{Phys. Rev. A} \textbf{\bibinfo{volume}{86}},
  \bibinfo{pages}{052334} (\bibinfo{year}{2012}).

\bibitem[{\citenamefont{Boixo et~al.}(2014)\citenamefont{Boixo, R{\o}nnow,
  Isakov, Wang, Wecker, Lidar, Martinis, and Troyer}}]{Boixo2014}
\bibinfo{author}{\bibfnamefont{S.}~\bibnamefont{Boixo}},
  \bibinfo{author}{\bibfnamefont{T.~F.} \bibnamefont{R{\o}nnow}},
  \bibinfo{author}{\bibfnamefont{S.~V.} \bibnamefont{Isakov}},
  \bibinfo{author}{\bibfnamefont{Z.}~\bibnamefont{Wang}},
  \bibinfo{author}{\bibfnamefont{D.}~\bibnamefont{Wecker}},
  \bibinfo{author}{\bibfnamefont{D.~A.} \bibnamefont{Lidar}},
  \bibinfo{author}{\bibfnamefont{J.~M.} \bibnamefont{Martinis}},
  \bibnamefont{and} \bibinfo{author}{\bibfnamefont{M.}~\bibnamefont{Troyer}},
  \bibinfo{journal}{Nature Physics} \textbf{\bibinfo{volume}{10}},
  \bibinfo{pages}{218} (\bibinfo{year}{2014}).

\bibitem[{\citenamefont{Katzgraber et~al.}(2014)\citenamefont{Katzgraber,
  Hamze, and Andrist}}]{Katzgraber2014}
\bibinfo{author}{\bibfnamefont{H.~G.} \bibnamefont{Katzgraber}},
  \bibinfo{author}{\bibfnamefont{F.}~\bibnamefont{Hamze}}, \bibnamefont{and}
  \bibinfo{author}{\bibfnamefont{R.~S.} \bibnamefont{Andrist}},
  \bibinfo{journal}{Phys. Rev. X} \textbf{\bibinfo{volume}{4}},
  \bibinfo{pages}{021008} (\bibinfo{year}{2014}).

\bibitem[{\citenamefont{Katzgraber et~al.}(2015)\citenamefont{Katzgraber,
  Hamze, Zhu, Ochoa, and Munoz-Bauza}}]{Katzgraber2015}
\bibinfo{author}{\bibfnamefont{H.~G.} \bibnamefont{Katzgraber}},
  \bibinfo{author}{\bibfnamefont{F.}~\bibnamefont{Hamze}},
  \bibinfo{author}{\bibfnamefont{Z.}~\bibnamefont{Zhu}},
  \bibinfo{author}{\bibfnamefont{A.~J.} \bibnamefont{Ochoa}}, \bibnamefont{and}
  \bibinfo{author}{\bibfnamefont{H.}~\bibnamefont{Munoz-Bauza}},
  \bibinfo{journal}{Phys. Rev. X} \textbf{\bibinfo{volume}{5}},
  \bibinfo{pages}{031026} (\bibinfo{year}{2015}).

\bibitem[{\citenamefont{Albash et~al.}(2015)\citenamefont{Albash, R{\o}nnow,
  Troyer, and Lidar}}]{Albash2015}
\bibinfo{author}{\bibfnamefont{T.}~\bibnamefont{Albash}},
  \bibinfo{author}{\bibfnamefont{T.}~\bibnamefont{R{\o}nnow}},
  \bibinfo{author}{\bibfnamefont{M.}~\bibnamefont{Troyer}}, \bibnamefont{and}
  \bibinfo{author}{\bibfnamefont{D.}~\bibnamefont{Lidar}},
  \bibinfo{journal}{Euro. Phys. J. Special Topics}
  \textbf{\bibinfo{volume}{224}}, \bibinfo{pages}{111} (\bibinfo{year}{2015}).

\bibitem[{\citenamefont{Heim et~al.}(2015)\citenamefont{Heim, R{\o}nnow,
  Isakov, and Troyer}}]{Heim2015}
\bibinfo{author}{\bibfnamefont{B.}~\bibnamefont{Heim}},
  \bibinfo{author}{\bibfnamefont{T.~F.} \bibnamefont{R{\o}nnow}},
  \bibinfo{author}{\bibfnamefont{S.~V.} \bibnamefont{Isakov}},
  \bibnamefont{and} \bibinfo{author}{\bibfnamefont{M.}~\bibnamefont{Troyer}},
  \bibinfo{journal}{Science} \textbf{\bibinfo{volume}{348}},
  \bibinfo{pages}{215} (\bibinfo{year}{2015}).

\bibitem[{\citenamefont{Hen et~al.}(2015)\citenamefont{Hen, Job, Job, Troyer,
  and Lidar}}]{Hen2015}
\bibinfo{author}{\bibfnamefont{I.}~\bibnamefont{Hen}},
  \bibinfo{author}{\bibfnamefont{J.}~\bibnamefont{Job}},
  \bibinfo{author}{\bibfnamefont{J.}~\bibnamefont{Job}},
  \bibinfo{author}{\bibfnamefont{M.}~\bibnamefont{Troyer}}, \bibnamefont{and}
  \bibinfo{author}{\bibfnamefont{D.~A.} \bibnamefont{Lidar}},
  \bibinfo{journal}{Phys. Rev. A} \textbf{\bibinfo{volume}{92}},
  \bibinfo{pages}{042325} (\bibinfo{year}{2015}).

\bibitem[{\citenamefont{{V. Isakov} et~al.}(2015)\citenamefont{{V. Isakov},
  Mazzola, {N. Smelyanskiy}, Jiang, Boixo, Neven, and Troyer}}]{Isakov2015}
\bibinfo{author}{\bibfnamefont{S.}~\bibnamefont{{V. Isakov}}},
  \bibinfo{author}{\bibfnamefont{G.}~\bibnamefont{Mazzola}},
  \bibinfo{author}{\bibfnamefont{V.}~\bibnamefont{{N. Smelyanskiy}}},
  \bibinfo{author}{\bibfnamefont{Z.}~\bibnamefont{Jiang}},
  \bibinfo{author}{\bibfnamefont{S.}~\bibnamefont{Boixo}},
  \bibinfo{author}{\bibfnamefont{H.}~\bibnamefont{Neven}}, \bibnamefont{and}
  \bibinfo{author}{\bibfnamefont{M.}~\bibnamefont{Troyer}}
  (\bibinfo{year}{2015}), \eprint{arXiv:1510.08057}.

\bibitem[{\citenamefont{Martin-Mayor and Hen}(2015)}]{Martin-mayor2015}
\bibinfo{author}{\bibfnamefont{V.}~\bibnamefont{Martin-Mayor}}
  \bibnamefont{and} \bibinfo{author}{\bibfnamefont{I.}~\bibnamefont{Hen}},
  \bibinfo{journal}{Sci. Rep.} \textbf{\bibinfo{volume}{5}},
  \bibinfo{pages}{15324} (\bibinfo{year}{2015}).

\bibitem[{\citenamefont{Liu et~al.}(2015)\citenamefont{Liu, Polkovnikov, and
  Sandvik}}]{Liu_2015}
\bibinfo{author}{\bibfnamefont{C.-W.} \bibnamefont{Liu}},
  \bibinfo{author}{\bibfnamefont{A.}~\bibnamefont{Polkovnikov}},
  \bibnamefont{and} \bibinfo{author}{\bibfnamefont{A.~W.}
  \bibnamefont{Sandvik}}, \bibinfo{journal}{Phys. Rev. Lett.}
  \textbf{\bibinfo{volume}{114}}, \bibinfo{pages}{147203}
  (\bibinfo{year}{2015}).

\bibitem[{\citenamefont{Steiger et~al.}(2015)\citenamefont{Steiger, R{\o}nnow,
  and Troyer}}]{Steiger2015}
\bibinfo{author}{\bibfnamefont{D.~S.} \bibnamefont{Steiger}},
  \bibinfo{author}{\bibfnamefont{T.~F.} \bibnamefont{R{\o}nnow}},
  \bibnamefont{and} \bibinfo{author}{\bibfnamefont{M.}~\bibnamefont{Troyer}},
  \bibinfo{journal}{Phys. Rev. Lett.} \textbf{\bibinfo{volume}{115}},
  \bibinfo{pages}{230501} (\bibinfo{year}{2015}).

\bibitem[{\citenamefont{Venturelli et~al.}(2015)\citenamefont{Venturelli,
  Mandr{\`{a}}, Knysh, O'Gorman, Biswas, and Smelyanskiy}}]{Venturelli2015}
\bibinfo{author}{\bibfnamefont{D.}~\bibnamefont{Venturelli}},
  \bibinfo{author}{\bibfnamefont{S.}~\bibnamefont{Mandr{\`{a}}}},
  \bibinfo{author}{\bibfnamefont{S.}~\bibnamefont{Knysh}},
  \bibinfo{author}{\bibfnamefont{B.}~\bibnamefont{O'Gorman}},
  \bibinfo{author}{\bibfnamefont{R.}~\bibnamefont{Biswas}}, \bibnamefont{and}
  \bibinfo{author}{\bibfnamefont{V.}~\bibnamefont{Smelyanskiy}},
  \bibinfo{journal}{Phys. Rev. X} \textbf{\bibinfo{volume}{5}},
  \bibinfo{pages}{031040} (\bibinfo{year}{2015}).

\bibitem[{\citenamefont{Crosson and Harrow}(2016)}]{Crosson2016}
\bibinfo{author}{\bibfnamefont{E.}~\bibnamefont{Crosson}} \bibnamefont{and}
  \bibinfo{author}{\bibfnamefont{A.~W.} \bibnamefont{Harrow}}
  (\bibinfo{year}{2016}), \eprint{arXiv:1601.03030}.

\bibitem[{\citenamefont{Denchev et~al.}(2016)\citenamefont{Denchev, Boixo,
  Isakov, Ding, Babbush, Smelyanskiy, Martinis, and Neven}}]{Denchev2016}
\bibinfo{author}{\bibfnamefont{V.~S.} \bibnamefont{Denchev}},
  \bibinfo{author}{\bibfnamefont{S.}~\bibnamefont{Boixo}},
  \bibinfo{author}{\bibfnamefont{S.~V.} \bibnamefont{Isakov}},
  \bibinfo{author}{\bibfnamefont{N.}~\bibnamefont{Ding}},
  \bibinfo{author}{\bibfnamefont{R.}~\bibnamefont{Babbush}},
  \bibinfo{author}{\bibfnamefont{V.}~\bibnamefont{Smelyanskiy}},
  \bibinfo{author}{\bibfnamefont{J.}~\bibnamefont{Martinis}}, \bibnamefont{and}
  \bibinfo{author}{\bibfnamefont{H.}~\bibnamefont{Neven}},
  \bibinfo{journal}{Phys. Rev. X} \textbf{\bibinfo{volume}{6}},
  \bibinfo{pages}{031015} (\bibinfo{year}{2016}).

\bibitem[{\citenamefont{Kechedzhi and Smelyanskiy}(2016)}]{Kechedzhi2016}
\bibinfo{author}{\bibfnamefont{K.}~\bibnamefont{Kechedzhi}} \bibnamefont{and}
  \bibinfo{author}{\bibfnamefont{V.~N.} \bibnamefont{Smelyanskiy}},
  \bibinfo{journal}{Phys. Rev. X} \textbf{\bibinfo{volume}{6}},
  \bibinfo{pages}{021028} (\bibinfo{year}{2016}).

\bibitem[{\citenamefont{Mandr{\`{a}}
  et~al.}(2016{\natexlab{a}})\citenamefont{Mandr{\`{a}}, Zhu, Wang,
  Perdomo-Ortiz, and Katzgraber}}]{Mandra2016}
\bibinfo{author}{\bibfnamefont{S.}~\bibnamefont{Mandr{\`{a}}}},
  \bibinfo{author}{\bibfnamefont{Z.}~\bibnamefont{Zhu}},
  \bibinfo{author}{\bibfnamefont{W.}~\bibnamefont{Wang}},
  \bibinfo{author}{\bibfnamefont{A.}~\bibnamefont{Perdomo-Ortiz}},
  \bibnamefont{and} \bibinfo{author}{\bibfnamefont{H.~G.}
  \bibnamefont{Katzgraber}}, \bibinfo{journal}{Phys. Rev. A}
  \textbf{\bibinfo{volume}{94}}, \bibinfo{pages}{022337}
  (\bibinfo{year}{2016}{\natexlab{a}}).

\bibitem[{\citenamefont{Mandr{\`{a}}
  et~al.}(2016{\natexlab{b}})\citenamefont{Mandr{\`{a}}, Zhu, and
  Katzgraber}}]{Mandra2016b}
\bibinfo{author}{\bibfnamefont{S.}~\bibnamefont{Mandr{\`{a}}}},
  \bibinfo{author}{\bibfnamefont{Z.}~\bibnamefont{Zhu}}, \bibnamefont{and}
  \bibinfo{author}{\bibfnamefont{H.~G.} \bibnamefont{Katzgraber}}
  (\bibinfo{year}{2016}{\natexlab{b}}), \eprint{arXiv:1606.07146}.

\bibitem[{\citenamefont{Marshall et~al.}(2016)\citenamefont{Marshall,
  Martin-Mayor, and Hen}}]{Marshall2016}
\bibinfo{author}{\bibfnamefont{J.}~\bibnamefont{Marshall}},
  \bibinfo{author}{\bibfnamefont{V.}~\bibnamefont{Martin-Mayor}},
  \bibnamefont{and} \bibinfo{author}{\bibfnamefont{I.}~\bibnamefont{Hen}},
  \bibinfo{journal}{Phys. Rev. A} \textbf{\bibinfo{volume}{94}},
  \bibinfo{pages}{012320} (\bibinfo{year}{2016}).

\bibitem[{\citenamefont{Muthukrishnan et~al.}(2016)\citenamefont{Muthukrishnan,
  Albash, and Lidar}}]{Muthukrishnan2016}
\bibinfo{author}{\bibfnamefont{S.}~\bibnamefont{Muthukrishnan}},
  \bibinfo{author}{\bibfnamefont{T.}~\bibnamefont{Albash}}, \bibnamefont{and}
  \bibinfo{author}{\bibfnamefont{D.~A.} \bibnamefont{Lidar}},
  \bibinfo{journal}{Phys. Rev. X} \textbf{\bibinfo{volume}{6}},
  \bibinfo{pages}{031010} (\bibinfo{year}{2016}).

\bibitem[{\citenamefont{Isakov et~al.}(2016)\citenamefont{Isakov, Mazzola,
  Smelyanskiy, Jiang, Boixo, Neven, and Troyer}}]{Isakov2016}
\bibinfo{author}{\bibfnamefont{S.~V.} \bibnamefont{Isakov}},
  \bibinfo{author}{\bibfnamefont{G.}~\bibnamefont{Mazzola}},
  \bibinfo{author}{\bibfnamefont{V.~N.} \bibnamefont{Smelyanskiy}},
  \bibinfo{author}{\bibfnamefont{Z.}~\bibnamefont{Jiang}},
  \bibinfo{author}{\bibfnamefont{S.}~\bibnamefont{Boixo}},
  \bibinfo{author}{\bibfnamefont{H.}~\bibnamefont{Neven}}, \bibnamefont{and}
  \bibinfo{author}{\bibfnamefont{M.}~\bibnamefont{Troyer}},
  \bibinfo{journal}{Phys. Rev. Lett.} \textbf{\bibinfo{volume}{117}},
  \bibinfo{pages}{180402} (\bibinfo{year}{2016}).

\bibitem[{\citenamefont{Albash and
  Lidar}(2017{\natexlab{a}})}]{AlbashLidar2016}
\bibinfo{author}{\bibfnamefont{T.}~\bibnamefont{Albash}} \bibnamefont{and}
  \bibinfo{author}{\bibfnamefont{D.~A.} \bibnamefont{Lidar}},
  \bibinfo{journal}{arXiv:1611.04471}  (\bibinfo{year}{2017}{\natexlab{a}}).

\bibitem[{\citenamefont{King et~al.}(2017)\citenamefont{King, Yarkoni, Raymond,
  Ozfidan, King, Nevisi, Hilton, and McGeoch}}]{King2017}
\bibinfo{author}{\bibfnamefont{J.}~\bibnamefont{King}},
  \bibinfo{author}{\bibfnamefont{S.}~\bibnamefont{Yarkoni}},
  \bibinfo{author}{\bibfnamefont{J.}~\bibnamefont{Raymond}},
  \bibinfo{author}{\bibfnamefont{I.}~\bibnamefont{Ozfidan}},
  \bibinfo{author}{\bibfnamefont{A.~D.} \bibnamefont{King}},
  \bibinfo{author}{\bibfnamefont{M.~M.} \bibnamefont{Nevisi}},
  \bibinfo{author}{\bibfnamefont{J.~P.} \bibnamefont{Hilton}},
  \bibnamefont{and} \bibinfo{author}{\bibfnamefont{C.~C.}
  \bibnamefont{McGeoch}}, \bibinfo{journal}{arXiv:1701.04579}
  (\bibinfo{year}{2017}).

\bibitem[{\citenamefont{Mandr{\`{a}} et~al.}(2017)\citenamefont{Mandr{\`{a}},
  Katzgraber, and Thomas}}]{Mandra2017}
\bibinfo{author}{\bibfnamefont{S.}~\bibnamefont{Mandr{\`{a}}}},
  \bibinfo{author}{\bibfnamefont{H.~G.} \bibnamefont{Katzgraber}},
  \bibnamefont{and} \bibinfo{author}{\bibfnamefont{C.}~\bibnamefont{Thomas}},
  \bibinfo{journal}{arXiv:1703.00622}  (\bibinfo{year}{2017}).

\bibitem[{\citenamefont{Albash and
  Lidar}(2017{\natexlab{b}})}]{AlbashLidar2017}
\bibinfo{author}{\bibfnamefont{T.}~\bibnamefont{Albash}} \bibnamefont{and}
  \bibinfo{author}{\bibfnamefont{D.~A.} \bibnamefont{Lidar}},
  \bibinfo{journal}{arXiv:1705.07452}  (\bibinfo{year}{2017}{\natexlab{b}}).

\bibitem[{\citenamefont{Jiang et~al.}(2017)\citenamefont{Jiang, Smelyanskiy,
  Isakov, Boixo, Mazzola, Troyer, and Neven}}]{Jiang2017}
\bibinfo{author}{\bibfnamefont{Z.}~\bibnamefont{Jiang}},
  \bibinfo{author}{\bibfnamefont{V.~N.} \bibnamefont{Smelyanskiy}},
  \bibinfo{author}{\bibfnamefont{S.~V.} \bibnamefont{Isakov}},
  \bibinfo{author}{\bibfnamefont{S.}~\bibnamefont{Boixo}},
  \bibinfo{author}{\bibfnamefont{G.}~\bibnamefont{Mazzola}},
  \bibinfo{author}{\bibfnamefont{M.}~\bibnamefont{Troyer}}, \bibnamefont{and}
  \bibinfo{author}{\bibfnamefont{H.}~\bibnamefont{Neven}},
  \bibinfo{journal}{Phys. Rev. A} p. \bibinfo{pages}{012322}
  (\bibinfo{year}{2017}).

\bibitem[{\citenamefont{Herr et~al.}(2017)\citenamefont{Herr, Brown, Heim,
  K{\"{o}}nz, Mazzola, and Troyer}}]{Herr2017}
\bibinfo{author}{\bibfnamefont{D.}~\bibnamefont{Herr}},
  \bibinfo{author}{\bibfnamefont{E.}~\bibnamefont{Brown}},
  \bibinfo{author}{\bibfnamefont{B.}~\bibnamefont{Heim}},
  \bibinfo{author}{\bibfnamefont{M.}~\bibnamefont{K{\"{o}}nz}},
  \bibinfo{author}{\bibfnamefont{G.}~\bibnamefont{Mazzola}}, \bibnamefont{and}
  \bibinfo{author}{\bibfnamefont{M.}~\bibnamefont{Troyer}},
  \bibinfo{journal}{arXiv:1704.00420}  (\bibinfo{year}{2017}).

\bibitem[{\citenamefont{Grover}(1997)}]{Grover_PRL97}
\bibinfo{author}{\bibfnamefont{L.~K.} \bibnamefont{Grover}},
  \bibinfo{journal}{Phys. Rev. Lett.} \textbf{\bibinfo{volume}{79}},
  \bibinfo{pages}{325} (\bibinfo{year}{1997}).

\bibitem[{\citenamefont{Roland and Cerf}(2002)}]{Roland_PRA02}
\bibinfo{author}{\bibfnamefont{J.}~\bibnamefont{Roland}} \bibnamefont{and}
  \bibinfo{author}{\bibfnamefont{N.~J.} \bibnamefont{Cerf}},
  \bibinfo{journal}{Phys. Rev. A} \textbf{\bibinfo{volume}{65}},
  \bibinfo{pages}{042308} (\bibinfo{year}{2002}).

\bibitem[{\citenamefont{J{\"o}rg et~al.}(2010)\citenamefont{J{\"o}rg, Krzakala,
  Kurchan, Maggs, and Pujos}}]{Jorg_EPL10}
\bibinfo{author}{\bibfnamefont{T.}~\bibnamefont{J{\"o}rg}},
  \bibinfo{author}{\bibfnamefont{F.}~\bibnamefont{Krzakala}},
  \bibinfo{author}{\bibfnamefont{J.}~\bibnamefont{Kurchan}},
  \bibinfo{author}{\bibfnamefont{A.~C.} \bibnamefont{Maggs}}, \bibnamefont{and}
  \bibinfo{author}{\bibfnamefont{J.}~\bibnamefont{Pujos}},
  \bibinfo{journal}{EPL} \textbf{\bibinfo{volume}{89}}, \bibinfo{pages}{40004}
  (\bibinfo{year}{2010}).

\bibitem[{\citenamefont{Caneva et~al.}(2008)\citenamefont{Caneva, Fazio, and
  Santoro}}]{Caneva_PRB08}
\bibinfo{author}{\bibfnamefont{T.}~\bibnamefont{Caneva}},
  \bibinfo{author}{\bibfnamefont{R.}~\bibnamefont{Fazio}}, \bibnamefont{and}
  \bibinfo{author}{\bibfnamefont{G.~E.} \bibnamefont{Santoro}},
  \bibinfo{journal}{Phys. Rev. B} \textbf{\bibinfo{volume}{78}},
  \bibinfo{pages}{104426} (\bibinfo{year}{2008}).

\bibitem[{\citenamefont{Bapst and Semerjian}(2012)}]{Bapst_JSTAT12}
\bibinfo{author}{\bibfnamefont{V.}~\bibnamefont{Bapst}} \bibnamefont{and}
  \bibinfo{author}{\bibfnamefont{G.}~\bibnamefont{Semerjian}},
  \bibinfo{journal}{JSTAT} p. \bibinfo{pages}{P06007} (\bibinfo{year}{2012}).

\bibitem[{\citenamefont{Nishimori and Takada}(2017)}]{NishimoriTakada2017}
\bibinfo{author}{\bibfnamefont{H.}~\bibnamefont{Nishimori}} \bibnamefont{and}
  \bibinfo{author}{\bibfnamefont{K.}~\bibnamefont{Takada}},
  \bibinfo{journal}{Frontiers in ICT} \textbf{\bibinfo{volume}{4}},
  \bibinfo{pages}{2} (\bibinfo{year}{2017}).

\bibitem[{\citenamefont{Seki and Nishimori}(2012)}]{Seki2012}
\bibinfo{author}{\bibfnamefont{Y.}~\bibnamefont{Seki}} \bibnamefont{and}
  \bibinfo{author}{\bibfnamefont{H.}~\bibnamefont{Nishimori}},
  \bibinfo{journal}{Phy. Rev. E} \textbf{\bibinfo{volume}{85}},
  \bibinfo{pages}{051112} (\bibinfo{year}{2012}).

\bibitem[{\citenamefont{Seoane and Nishimori}(2012)}]{Seoane2012}
\bibinfo{author}{\bibfnamefont{B.}~\bibnamefont{Seoane}} \bibnamefont{and}
  \bibinfo{author}{\bibfnamefont{H.}~\bibnamefont{Nishimori}},
  \bibinfo{journal}{J. Phys. A: Math. Theor.} \textbf{\bibinfo{volume}{45}},
  \bibinfo{pages}{435301} (\bibinfo{year}{2012}).

\bibitem[{\citenamefont{Susa et~al.}(2017)\citenamefont{Susa, Jadebeck, and
  Nishimori}}]{Susa2017}
\bibinfo{author}{\bibfnamefont{Y.}~\bibnamefont{Susa}},
  \bibinfo{author}{\bibfnamefont{J.~F.} \bibnamefont{Jadebeck}},
  \bibnamefont{and}
  \bibinfo{author}{\bibfnamefont{H.}~\bibnamefont{Nishimori}},
  \bibinfo{journal}{Physical Review A} \textbf{\bibinfo{volume}{95}},
  \bibinfo{pages}{042321} (\bibinfo{year}{2017}).

\bibitem[{\citenamefont{Nishimori et~al.}(2015)\citenamefont{Nishimori, Tsuda,
  and Knysh}}]{Nishimori_PRE14}
\bibinfo{author}{\bibfnamefont{H.}~\bibnamefont{Nishimori}},
  \bibinfo{author}{\bibfnamefont{J.}~\bibnamefont{Tsuda}}, \bibnamefont{and}
  \bibinfo{author}{\bibfnamefont{S.}~\bibnamefont{Knysh}},
  \bibinfo{journal}{Phys. Rev. E} \textbf{\bibinfo{volume}{91}},
  \bibinfo{pages}{012104} (\bibinfo{year}{2015}).

\bibitem[{\citenamefont{Zanca and Santoro}(2016)}]{Zanca_2016}
\bibinfo{author}{\bibfnamefont{T.}~\bibnamefont{Zanca}} \bibnamefont{and}
  \bibinfo{author}{\bibfnamefont{G.~E.} \bibnamefont{Santoro}},
  \bibinfo{journal}{Phys. Rev. B} \textbf{\bibinfo{volume}{93}},
  \bibinfo{pages}{224431} (\bibinfo{year}{2016}).

\bibitem[{\citenamefont{Glauber}(1963)}]{Glauber_JMP63}
\bibinfo{author}{\bibfnamefont{R.~J.} \bibnamefont{Glauber}},
  \bibinfo{journal}{J. Math. Phys.} \textbf{\bibinfo{volume}{4}},
  \bibinfo{pages}{294} (\bibinfo{year}{1963}).

\bibitem[{\citenamefont{van Kampen}(1992)}]{vanKampen:book}
\bibinfo{author}{\bibfnamefont{N.~G.} \bibnamefont{van Kampen}},
  \emph{\bibinfo{title}{Stochastic processes in physics and chemistry}}
  (\bibinfo{publisher}{North-Holland}, \bibinfo{year}{1992}),
  \bibinfo{edition}{{Revised and enlarged}} ed.

\bibitem[{\citenamefont{Suzuki and Okada}(2005)}]{Suzuki_JPSJ05}
\bibinfo{author}{\bibfnamefont{S.}~\bibnamefont{Suzuki}} \bibnamefont{and}
  \bibinfo{author}{\bibfnamefont{M.}~\bibnamefont{Okada}}, \bibinfo{journal}{J.
  Phys. Soc. Jpn.} \textbf{\bibinfo{volume}{74}}, \bibinfo{pages}{1649}
  (\bibinfo{year}{2005}).

\bibitem[{\citenamefont{De~Grandi et~al.}(2011)\citenamefont{De~Grandi,
  Polkovnikov, and Sandvik}}]{DeGrandi_PRB11}
\bibinfo{author}{\bibfnamefont{C.}~\bibnamefont{De~Grandi}},
  \bibinfo{author}{\bibfnamefont{A.}~\bibnamefont{Polkovnikov}},
  \bibnamefont{and} \bibinfo{author}{\bibfnamefont{A.~W.}
  \bibnamefont{Sandvik}}, \bibinfo{journal}{Phys. Rev. B}
  \textbf{\bibinfo{volume}{84}}, \bibinfo{pages}{224303}
  (\bibinfo{year}{2011}).

\end{thebibliography}

\end{document}